# Coarse-Grid Computational Fluid Dynamic (CG-CFD) Error Prediction using Machine Learning

**Botros N Hanna**
Department of Nuclear Engineering,
North Carolina State University,
Raleigh, NC, USA 27695-7909
e-mail: bnhannab@ncsu.edu

**Nam T. Dinh**
Professor
Department of Nuclear Engineering,
North Carolina State University,
Raleigh, NC, USA 27695-7909
e-mail: ntdinh@ncsu.edu
ASME Member

**Robert W. Youngblood**
Adjunct Professor
Idaho National Laboratory
INL, P.O. Box 1625,
Idaho Falls, ID 83415-3870, USA
e-mail: robert.youngblood@inl.gov

**Igor A. Bolotnov**
Associate Professor
Department of Nuclear Engineering,
North Carolina State University,
Raleigh, NC, USA 27695-7909
e-mail: iabolotn@ncsu.edu

## ABSTRACT

*Despite the progress in high performance computing, Computational Fluid Dynamics (CFD) simulations are*

*still computationally expensive for many practical engineering applications such as simulating large*

*computational domains and highly turbulent flows. One of the major reasons of the high expense of CFD is*

*the need for a fine grid to resolve phenomena at the relevant scale, and obtain a grid-independent solution. The fine grid requirements often drive the computational time step size down, which makes long transient problems prohibitively expensive. In the research presented, the feasibility of a Coarse Grid CFD (CG-CFD) approach is investigated by utilizing Machine Learning (ML) algorithms. Relying on coarse grids increases the discretization error. Hence, a method is suggested to produce a surrogate model that predicts the CG-CFD local errors to correct the variables of interest. Given high-fidelity data, a surrogate model is trained to predict the CG-CFD local errors as a function of the coarse grid local features. ML regression algorithms are utilized to construct a surrogate model that relates the local error and the coarse grid features. This method is applied to a three-dimensional flow in a lid driven cubic cavity domain. The performance of the method was assessed by training the surrogate model on the flow full field spatial data and tested on new data (from flows of different Reynolds number and/or computed by different grid sizes). The proposed method maximizes the benefit of the available data and shows potential for a good predictive capability.*

*__Keywords__: coarse grid (mesh), CFD, machine learning, discretization error, big data, artificial neural network, random forest regression, data-driven.*

## 1. INTRODUCTION

### 1.1. Motivation: The Need for CG-CFD

In performing high-fidelity simulations of complex fluid flows, Computational Fluid Dynamics (CFD) approach has an advantage over traditional physical modeling, because of its capability to provide detailed information about flow field. However, CFD is too computationally expensive for many real-world applications such as oceanic flows, nuclear energy applications, and indoor environment research (discussed below).

*a)*      *Oceanic Flows*

An example of this computational challenge is simulating the turbulent ocean currents in the North Atlantic that contain circulations with length scales ranging from few meters to thousands of kilometers [1]. Capturing the full range of the circulations' length scales using Direct Numerical Simulation (DNS) [2] requires very high-resolution grids and thus large computational power. On the other side, when modeling the average effect of turbulence on ocean currents, turbulence models for ocean simulations remove fluctuations of scales smaller than about 30 kilometers [1]. Hence, the energy in these "small"–scale fluctuations are lost, so the turbulence models are inaccurate.

*b)*      *Nuclear Energy Applications*

CFD's computational expense problem also arises in nuclear engineering applications. For instance, CFD simulation of a full reactor core, with its fuel rods, spacer grids, and other components, is not yet practical [3]. Utilizing CFD in nuclear reactor accident analyses is also computationally overwhelming, mainly due to long-time transient problems. For example, CFD was used to analyze only one localized phenomenon that may occur during nuclear accidents: high-pressure blowdown flow from the reactor cooling system needed over a week of computing time on a 128-processor cluster to simulate ten seconds of steam blowdown on a relatively coarse computational grid [4]. It would be computationally prohibitive to simulate complex, long transient and accident scenarios in this way.

*c)        Indoor Environments*

In turbulence modeling of indoor environments, research has been done aiming to perform real-time simulation for cases like emergency evacuation of buildings [5]. Real time simulation is not possible with the grid independent CFD. Hence, research was performed to speed up CFD simulations by using coarse grids [5-7].

## 1.2.    The Need for Data-Driven Models

In CFD, high-fidelity results can be obtained by simulating all the turbulence length and time scales using DNS [2], which is accurate but computationally challenging for many applications (turbulent flows with high Reynolds number). A computationally cheaper alternative approach is Large Eddy Simulation (LES) [8], that requires a computational grid that is coarser than the grid needed in DNS. In LES, the large-scale motions (energy-containing eddies) are captured, while the small-scale flow motions are either modelled using an explicit sub-grid scale model, or implicitly modeled using the numerical dissipation associated with the computational scheme knows as Implicit LES (ILES) (see [9, 10]).

Despite the progress in LES methods, the Reynolds Averaged Navier Stokes (RANS) equations [11] approach is still more popular and computationally economical in industrial applications. RANS approach requires a coarse grid (compared to LES and DNS) as it resolves the mean flow variables only, not the detailed instantaneous flow. However, RANS models (such as $k - \varepsilon$ [12] and Spalart-Allmaras [13] model) lack adaptability, i.e. RANS models perform well only for specific flow conditions and geometries. This gives an advantage to data-driven surrogate statistical models as they could adapt via data assimilation as more data becomes available. The availability of high-fidelity simulations provides an opportunity to inform data-driven coarse grid models. Currently, as discussed below, several groups of researchers are pursuing the development of data-driven methods to perform coarse grid simulations (see Section 1.3).

### 1.3. Progress in Data-Driven Models in Fluid Dynamics

Traditionally, fluid flow models (for turbulence or multiphase flow, etc.) were developed based on physical understanding of the phenomena often with certain empirical assumptions. Recently, data-driven models have been developed based on data processing and analysis (data-driven), in order to make more use of the available data. An overview of recent data-driven models is presented below. These data-driven models are either utilized in the context of CFD (to correct LES or RANS simulations or compute closure terms for averaged variables' equations) or in the context of Smoother Particle Hydrodynamics (SPH) [14].

*a)     Autonomic sub-grid Large Eddy Simulation (ALES)*

Autonomic Sub-grid Large Eddy Simulation (ALES) [15] can be regarded as a data-driven method that utilizes a grid that is deemed to be coarse compared to DNS grid requirements, but must be relatively fine to resolve Turbulence Kinetic Energy (TKE) in the inertial range of the turbulence energy spectrum. The ALES method expresses the local sub-grid scale stress tensor as a non-linear function of the resolved variables at all locations and all times with a Volterra series (which is similar to a Taylor series). The series coefficients are computed by minimizing the error in sub-grid scale stresses at a test filter scale. Then, coefficients are mapped to the LES scale assuming scale similarity (noting that both LES scale and test scale lie in the self-similar inertial range of the turbulence energy spectrum). The ALES method is a general model-free self-optimizing approach for closure of turbulence simulations. The ALES method needs neither previous training based on DNS results nor user specified parameters. Preliminary results with simple problems (homogenous turbulence and sheared turbulence) showed high accuracy in computing turbulent stresses compared to current turbulence models [16]. However, the scale similarity assumption, upon which ALES is based, is valid only in the inertial range (in turbulent kinetic energy spectrum). Therefore, applying ALES approach is still impractical in simulating large domains / long transients.

b)      *Closure Term in a Bubbly Flow Equation*

Another example [17] from the multiphase flow field is finding the closure term in an averaged simple equation for bubbly flow using an Artificial Neural Network (ANN) [18], given accurate results obtained by DNS. In that problem, the initial vertical velocity and the average bubble density are uniform except in one of the horizontal directions. As the transient develops, the bubble density and velocity become uniform. It was assumed that the unknown closure term depends on the void fraction, the void fraction gradient, and the liquid velocity gradient. The ANN "learns" the correlation between the closure term and these three variables from one simulation dataset. A transient of different initial conditions can then be predicted using that correlation. That case study was quite simple (the averaged equation is one-dimensional and the boundary conditions are periodic).

c)      *RANS Model Discrepancy*

To reduce the RANS model discrepancy by learning from data, research groups from University of Michigan and Virginia Tech University utilized ML techniques to predict or reduce the error in RANS simulation results [19, 20]. Both groups made use of high-fidelity simulation results from DNS to correct the low-fidelity model results from RANS. In [19], RANS results are corrected by spatially distributed multiplicative discrepancy term, $\beta$, into one of the terms of the transport equation of turbulent quantities; in [20], the Reynolds stress discrepancy is directly based on mean flow features computed by RANS. The statistical model in both cases is trained based on the available high-fidelity data and then tested on other cases, which have a different Reynolds number or a different flow geometry. Recently, the Virginia Tech University research group suggested [21] that the group of input features could be reduced to 4 flow features. It was also shown that not only could Reynolds stress discrepancy computed by RANS be improved by predicting Reynolds stress discrepancy, but also the improved Reynolds stress results in a more accurate velocity field.

Statistical data-driven models are usually trained on a set of data and then tested with another group of data that are "close" to the training data in some sense. The

"closeness" between training data and testing data can be used to assess the prediction

confidence *a priori* [22, 23]. This closeness could be quantified by metrics like the

Mahalanobis Distance [24] or the Kernel Density Estimation technique [25]. This approach was successful in most cases except when the training and testing cases are too close or too far apart [23].

ML model predictions, for RANS Reynolds stress anisotropy, for flows of similar physics,

should be similar. ML model predictions should not change with changing the

orientation of coordinate frame. In other words, ML predictions should be Galilean

invariant. Based on this principle, a multi-layer ANN (also called "deep learning"), with

invariant tensor basis, was proposed [26]. This way, Reynolds stress anisotropy tensor

can be predicted with embedded Galilean invariance. The proposed tensor-basis ANN

proved to have more accurate prediction compared to both RANS models and conventional ANN.

d)      *Smoothed Particle Hydrodynamics*

Data-driven models, for fluid simulations, are used not only in CFD, but also in

Smoothed Particle Hydrodynamics (SPH) (where Navier Stokes (NS) equations are

approximated on fluid particles instead of a computational grid) [14].Random Forest

Regression (RFR) [27] was trained to predict the velocity and position of the fluid particle in the next time step based on the velocity and the position in the previous time step [28]. This approach aims to "learn" the behavior of fluid from the training examples and provides an alternative for real time fluid simulations.

### 1.4. Scope of The Present Work

Since the traditional high-resolution (mesh-independent) CFD solution is computationally expensive and using of coarse grid would result in high grid-induced errors, the present work focuses on predicting the errors of the coarse grid solution. While the previous research aimed largely to reduce the model form error, in this work, the sub-grid effect is compensated for by a surrogate statistical model that is trained by using high- fidelity simulation results (fine-grid CFD).

The objective of this work is to investigate the feasibility of obtaining a correction for CG-CFD simulation results using ML algorithms. Numerical experiments are designed and performed to study the feasibility of utilizing ML tools to get a correlation between the 'correct' solution informed by fine grid simulation results and the coarse grid variables.

Among the different sources of error in CFD simulation, the turbulence modeling error and the discretization error are the most challenging ones. Below, in Subsections (a) and (b), the present work is compared to other CFD approaches in terms of mitigating turbulence modeling error and discretization error.

*a)    Present Approach vs. Traditional CFD*

RANS turbulence models are popular because of the reduced computational expense while representing three-dimensional flow behavior. RANS turbulence models typically rely on incorporating more physics and using empirical models for some parameters based on the available validation data. It is always required to get grid-independent solution (with a grid convergence study), but this grid convergence study should be applied within the range of grids that satisfy the turbulence model grid requirements. Typically, for each new case, a new simulation is needed, even if the new case is only slightly different from old cases

In the present work, the "No model" approach (where NS equations are solved numerically without any turbulence model) is utilized with coarse grids (discretization grids that are expected to produce non-accurate results). However, instead of investigating the turbulence modeling error, we use the same "No model" approach with different grids (fine and coarse grids) to train a surrogate model to compute this grid-induced error. CG-CFD error is predicted by comparing high-fidelity results (with fine grids) against CG-CFD results. With this surrogate model, the grid-induced error can be predicted for other cases that have different grid sizes, Reynolds numbers, etc. The surrogate model is adaptive with the available data so each new set of experimental data or high-fidelity computations will be reflected in the automatically-improved model, without the need for developing other models that capture the new data.

b)      *Present Approach vs. Physics Informed Machine Learning*

Recently, research has been pursued in the direction of taking advantage of the ML algorithms to develop surrogate models that can compute the turbulence modeling error (for instance, [19, 20, 26]). This approach is known as Physics Informed Machine Learning (PIML) approach. Both the present work and PIML method benefit from ML algorithms to produce data-driven statistical models. However, PIML assumes that one of the RANS models is used and the grid size criterion for this model is satisfied. Through PIML, the difference between RANS flow variables profile and DNS profile is computed. On the other side, the present work depends on the same Navier-Stokes equations used with both fine and coarse grids. After that, the error resulting from the grid coarseness is computed with the ML surrogate model.

### 1.5. The Structure of This Paper

In this paper, a method is proposed to predict the CG-CFD induced error given coarse grid features using ML surrogate models. This method is presented in Section 2. For purposes of comparison, two ML algorithms (ANN [18] and Random Forest Regression (RFR) [27] in MATLAB [29]) are each used to construct a surrogate model; these surrogate models are presented in Section 3. This method was applied to a three-dimensional quasi steady state turbulent flow inside a lid-driven cavity. The capability of the statistical model to predict the coarse grid-induced error in different cases (different Reynolds number and grid sizes) is illustrated and assessed in Section 4. The conclusions of this paper are provided in Section 5.

### 2.    METHOD

The problem statement is presented in Section 2.1 and depicted in Fig. 1. CG-CFD error can be predicted under some hypothesis (see Section 2.2). The proposed method for computing CG-CFD grid-induced error is explained in Section 2.3 and depicted in Fig. 2. CG-CFD approach is applied and tested with a three-dimensional flow in a lid-driven cubic cavity (see Section 2.4 and Fig. 3).

### 2.1. Problem Statement

The general problem to be addressed in this work can be formulated as illustrated in Fig. 1. In Fig. 1, the vertical axis $\psi$ is a number or a set of numbers that characterize the flow pattern (such as Reynolds number, Rayleigh number, domain aspect ratio, etc.), while the horizontal axis is the computational grid spacing, $\Delta$. For a specific problem of interest (in this work it is the flow inside a lid-driven cubic cavity), the green-colored area on the figure represents the high-fidelity data computed by sufficiently fine grids ($\Delta$ approaching zero). The subscript $f$ refers to a fine grid and $\varphi_f$ is the flow variable of interest (e.g. velocity component or pressure) computed by sufficiently fine grid. The red-colored area refers to low-fidelity computations, of $\varphi$, performed by coarse grids, that may or may not have corresponding available high-fidelity computations, $\varphi_f$. The problem here is how to use the low-fidelity and corresponding high-fidelity data to construct a function that can predict the correct value of a new (not available) $\varphi$, given only the corresponding low-fidelity computation only.

Traditionally, a new high-fidelity computationally expensive simulation is performed for each new $\psi$. In this work, the benefit of the data is maximized using the available high-fidelity data to capture the relation between coarse and fine grid data to predict $\varphi$ for a new case.

## 2.2. Research Hypothesis

The present hypothesis is that the CG-CFD error is predictable, using data-driven surrogate model, under the following assumptions:

▪ The training flows (to train the surrogate model) and the testing flows (to test the trained surrogate model) have similar physics with similar complexity. In order to correct the grid-induced error over the whole domain (for the training flows), high-fidelity data over the whole domain are needed.

▪ When comparing the fine and coarse grid data, the number of the cells in both grids is not the same. Thus, to compute local grid error, it is necessary to perform mapping of the fine grid data onto the coarse grid. In other words, $\varphi_f$, is replaced by $\varphi_{f \to \Delta}$ (the fine grid field of $\varphi$ mapped on a grid whose cell length is $\Delta$). This mapping (averaging) constitutes a source of error because of losing some details of the flow field profile. However, this mapped field is much more accurate than the field computed by a coarse grid (for instance, see Fig. 4).

▪ Using a coarse grid in the CG-CFD framework is limited by the scale of the phenomena of interest. If the phenomena of interest have a length scale, $l$, the coarse grids used should have a grid spacing, $\Delta < l$. It is assumed that the phenomena of interest have a scale that is large enough that a data-driven surrogate model can be used to predict the grid-induced error occurred due to the inability of the coarse grid to resolve sub-grid length-scales.

▪ Sub-grid flow features, computed by a coarse grid, are expected to be inaccurate. However, it is hypothesized that the grid-induced error is a function of the inaccurate coarse grid features.

▪ For most of the similar efforts in the literature (e.g. [19, 20, 23]) the aim was to get a data-driven surrogate model that compensates for sub-grid effects; the ultimate goal was predicting a quantity of interest (such as the lift force or Nusselt number) or specific linear profile (such as the axial velocity profiles). On the other hand, in the present work, it was assumed that we are interested in *all* the data through the whole domain (flow variable value at each grid cell). Thus, the proposed approach makes more use of the full field data.

### 2.3.  Proposed Approach

This section presents the proposed method for CG-CFD error prediction using ML algorithms. In the following paragraphs, we will discuss the error prediction method, feature selection, and ML error evaluation.

*a)  CG-CFD Error Prediction Method*

The proposed method illustrated in Fig. 2 consists of 2 sets of flows: training flows and testing flows. Training flows are utilized to construct a surrogate model of the grid-induced error as a function of the coarse grid flow features. High-fidelity simulations of the training flows are performed with sufficiently fine grids and represented by $\varphi_f^{tr}$ ($tr$ refers to training flows). Next, the same simulation is performed again with a coarse grid producing $\varphi_\Delta^{tr}$. $\varphi_f^{tr}$ is mapped onto the coarse grid, $\Delta$, to obtain $\varphi_{f\to\Delta}^{tr}$. Based on the coarse grid simulations of the training flows, flow "features", $\boldsymbol{X}(\varphi_\Delta^{tr})$, are computed. These features are the inputs to a surrogate model whose output is the local grid induced error, $\varepsilon_\Delta(\varphi^{tr})$:

$$\varepsilon_\Delta(\varphi^{tr}) = \varphi_{f\to\Delta}^{tr} - \varphi_\Delta^{tr} = F(\boldsymbol{X}(\varphi_\Delta)) \tag{1}$$

Selection of the features is discussed in the next section. The function $F$ is obtained from ML algorithms (ANN and RFR). The capability of the surrogate model, obtained from equation (1), is assessed by applying the model on testing flows as illustrated in Fig. 2.

### b)    Features Selection

Assuming a smooth profile for the flow variable, $\varphi$, consider a Taylor series expansion for a flow field variable $\varphi$ along the $x$-direction in a grid with cell length of $\Delta x$:

$$\varphi = \varphi_0 + \Delta x \left.\frac{d\varphi}{dx}\right|_{\Delta x} + \frac{(\Delta x)^2}{2} \left.\frac{d^2\varphi}{dx^2}\right|_{\Delta x} + \cdots \tag{2}$$

For a linear state variable profile, the grid-induced error will be of order $(\Delta x)^2 \left.\frac{d^2\varphi}{dx^2}\right|_{\Delta x}$ and the effect of $\Delta x \left.\frac{d^2\varphi}{dx^2}\right|_{\Delta x}$ may be considered for non-linear profiles. The flow state variables' derivatives were utilized before as indicators of the need for high local refinement in the adaptive-grid-refinement literature [30-32]. This was inspired by the general observation that a finer grid is needed near a discontinuity or steep curve of the solution. Another idea to be considered is the fact that the value of the variable $\varphi$ and its corresponding error at any grid cell is dependent on the value of $\varphi$ and the corresponding error at the neighboring cells. Hence, the derivatives $\frac{d\varphi}{dx}$ and $\frac{d^2\varphi}{dx^2}$ are carrying the effect of the neighboring cells.

Additionally, the fluid flow is typically characterized by global numbers like Reynolds number, Rayleigh number, etc. Thus, we propose to utilize a local number (as flow feature) which substitutes for the global number. In this work, the quantity of interest is the velocity. Therefore, we selected the cell Reynolds number as the local quantity of interest. The cell Reynolds number substitutes for the global velocity and length scales with local velocity and cell length, $\Delta$. The cell Reynolds number, $Re_\Delta$, defined in terms of the local velocity magnitude, $|U|$ and the kinematic viscosity, $\nu$, is given by:

$$Re_\Delta = |U|\Delta/\nu \tag{3}$$

Based on the previous discussions, the proposed features vector, $X(\varphi_\Delta)$, becomes:

$$X(\varphi_\Delta) = \left( Re_\Delta, \Delta x_j \frac{du_i}{dx_j}\bigg|_{\Delta x_j}, (\Delta x_j)^2 \frac{d^2u_i}{dx_j{}^2}\bigg|_{\Delta x_j} \right) \tag{4}$$

Note that the term $\Delta x_j \frac{du_i}{dx_j}$ represents 9 features because the velocity has 3 components $(U_x, U_y, U_z)$ and the derivative is performed with respect to 3 directions $(x, y, z)$ (for instance: $\frac{dU_x}{dx}, \frac{dU_x}{dy}, etc.$ Similarly, the term $(\Delta x_j)^2 \frac{d^2u_i}{dx_j{}^2}$ represents 27 features (accounting for all the second derivatives for the 3 velocity components).

When comparing different sets of CFD data, it is desired to use dimensionless data for performing comparisons between different geometries; hence, the features are normalized using the length scale and the velocity scale (see Table 1). The normalized features, $\widetilde{X}$, and the normalized grid induced-error, $\tilde{\varepsilon}_{\Delta}^{tr}$, are incorporated into equation (1) as:

$$\tilde{\varepsilon}_{\Delta}(\varphi^{tr}) = F\left(\widetilde{X}(\varphi_{\Delta})\right) \tag{5}$$

In this work (a lid-driven cavity flow), the length scale, $H$, is the cavity height, and the velocity scale is the cavity lid velocity, $U_{lid}$.

c)    *Machine Learning Error Assessment*

To assess the accuracy of the surrogate model produced by ML algorithm, a metric, $E_{ML}$ is proposed. This metric (ML error) is calculated as:

$$E_{ML} = \frac{|\tilde{\varepsilon}_{\Delta}^{act}(\varphi) - \tilde{\varepsilon}_{\Delta}^{ML}(\varphi)|}{\sigma_{\tilde{\varepsilon}_{\Delta}^{act}(\varphi)}} \tag{6}$$

where $\tilde{\varepsilon}_{\Delta}^{act}(\varphi)$ is the actual grid-induced error when computing the variable $\varphi$ and $\tilde{\varepsilon}_{\Delta}^{ML}(\varphi)$ is the estimated error. The difference between the actual and ML prediction of the grid-induced error is normalized by the standard deviation of the actual grid-induced error, $\sigma_{\tilde{\varepsilon}_{\Delta}^{act}(\varphi)}$. Typically, when computing the surrogate model error, the difference between the actual and estimated value is divided by the actual value to get a relative error. However, here, we divide by the standard deviation of the actual value to avoid having zeros in the denominator. Additionally, computing ML error in Eq. (6) gives the ML error in terms of the standard deviation of the grid-induced error.

## 2.4.    Case Study

The lid-driven cavity flow is one of the most studied problems in CFD [33, 34]. Although the problem can be stated relatively simply, the flow in a cavity features interesting flow physics with counter-rotating vortices appearing at the corners of the cavity [34], and it serves as a benchmark problem for numerical methods [35-37].

The case studied here is a fluid contained in a cube whose height is one meter, with five fixed walls and one lid, moving with velocity, $U_{lid} = 1 \text{ m/s}$ (see Fig. 3). The physics of this problem is related to the global Reynolds number, $Re$, which is defined in terms of the velocity magnitude, $|U|$, the fluid kinematic viscosity, $\nu$, and the cavity characteristic length (height), $H$.

$$Re = \frac{|U|H}{\nu} \qquad (7)$$

At a Reynolds number between 2000 and 3000, an instability appears in the vicinity of the downstream corner [38]. As expected, the turbulence level near the wall increases with the increasing Reynolds number. The flow becomes fully turbulent near the downstream corner eddy at Reynolds number higher than 10,000 [38].

### a)    Governing Equations

In this work, CFD simulations were performed with one of the available open source CFD software packages, *OpenFOAM* [39]. All the simulations were performed with three-dimensional NS incompressible flow of a viscous Newtonian fluid [2]:

$$\frac{\partial u_i}{\partial x_i} = 0 \tag{8}$$

$$\frac{\partial u_i}{\partial t} + \frac{\partial u_i u_j}{\partial x_j} = \nu \frac{\partial^2 u_i}{\partial x_j \partial x_j} - \frac{\partial p}{\partial x_i} \tag{9}$$

where $u_i$ is the velocity component $i$, $p$ is the kinematic pressure and $\nu$ is the kinematic viscosity. NS equations are discretized in space using finite volume method on meshes composed of hexahedral cells. NS equations are integrated over the volume of the computational cells [40]. The time dependent term is assumed constant over the computational cell volume. The divergence theorem is applied to convection, pressure gradient and diffusion terms:

$$\iint_A dA(u_i n_i) = 0 \tag{10}$$

$$V \frac{\partial u_i}{\partial t} + \iint dA u_i u_j n_j = \iint dA \nu \frac{\partial u_i}{\partial x_j} n_j - \iint dA p n_i \tag{11}$$

Where $V$ is the computational cell volume, $n$ is the normal of the surface of the control volume and $A$ is the surface area. The area integrals can be written as summations over each face:

$$\sum_{nbr} (u_i n_i A)_{nbr} = 0 \tag{12}$$

$$V \frac{\partial u_i}{\partial t} + \sum_{nbr} (u_i u_j n_j A)_{nbr} = \sum_{nbr} \left( \nu \frac{\partial u_i}{\partial x_j} n_j A \right)_{nbr} - \sum_{nbr} (p n_i A)_{nbr} \tag{13}$$

The subscript $nbr$ refers to the value at any given face. The time derivative term is approximated using a second order scheme (backward difference). For the spatial derivatives, the central difference scheme (second order accurate) has been used [41].

*b)    Numerical procedures*

High-fidelity Simulations were performed on a $120 \times 120 \times 120$ grid, with special refinement towards the wall. The length of the cells touching the wall is $0.0014$ meters so the total number of cells in the fine grid is $2 \times 10^6$ cells.

In OpenFOAM, a transient solver (named pisoFoam) for incompressible, turbulent flow, using the PISO algorithm [42], was used. Using pisoFoam, unsteady simulation can be performed, but the focus of the present work is the quasi steady-state three-dimensional flow inside the cavity.

*c)    Training and Testing Cases*

When training a surrogate data-driven model, the predictive capability of the model is impacted by the similarity or the difference between the training flows and the testing flow. Thus, to assess the surrogate model, different scenarios of the available training data and the targeted testing case are studied (see the list of scenarios in Table 2).

**3.    MACHINE LEARNING**

It is a new challenge to explore/ analyze/ model/ exploit the "big data" generated by high-resolution numerical simulation and physical experiments. ML algorithms can be used to find patterns in data and build new models that predict new outcomes based on historical data. In general, ML algorithms are classified depending on the available information and the user purpose in two types: supervised learning and unsupervised learning. Supervised learning demands having a set of input variables (also called features) and their corresponding outputs. These inputs and outputs help to train a model that could predict the output for a new input. Supervised learning includes applications like regression and classification. On the other side, unsupervised learning is more challenging because the observations (raw data) are just given without identifying inputs or outputs. Its function is to extract the internal relationships in the dataset. Examples of unsupervised learning are: grouping the data (clustering), or recognizing anomalies in the dataset, or discovering the association between some variables. In the present work, the ML algorithms selected are ANN and RFR. These techniques are used in this work for regression, and their effectiveness is compared. ANN and RFR are presented in Sections 3.1 and 3.2.

## 3.1    Artificial Neural Network (ANN)

Artificial Neural Network (ANN) is a well-known non-linear adaptive regression tool that identifies a function that relates the given inputs and outputs. ANN structure was originally intended to mimic that of human brain. ANN consists of computational units that are named artificial neurons. The biological neuron receives signals and activates if the signal is strong (exceeds a threshold), and then sends the signal to another neuron, and so on. The artificial neuron executes a similar process: it receives input variables which are multiplied by weights (which modify the signal strength), and are then processed by a mathematical activation function (like neuron activation) to produce the output.

a)      *ANN Structure*

ANN structure can be represented mathematically as follows:

$$\underset{S\times 1}{\boldsymbol{a}} = f(\underset{S\times R}{\boldsymbol{w}}\ \underset{R\times 1}{\boldsymbol{X}} + \underset{S\times 1}{\boldsymbol{b}}) \tag{14}$$

$$\underset{1\times 1}{Y} = \underset{1\times S}{\boldsymbol{\alpha}}\ \underset{S\times 1}{\boldsymbol{a}} + \underset{1\times 1}{\lambda} \tag{15}$$

where $X$ is the vector of features (inputs) and $Y$ is the output. The number of inputs is $R$ and the number of neurons is $S$. $f$ is the activation function, which is most often chosen to be a hyperbolic tangent function. For instance, in our case, $X$ is a vector of flow features at any point on the coarse grid of the simulation domain and the output variable $Y$ is the local coarse grid-induced error. Given some number of samples, the weights, $w$ and $\alpha$, and the biases, $b$ and $\lambda$, are adjusted to optimize the performance of the ANN by minimizing the Mean Squared Error (MSE). The MSE at each sample (data point) is defined as follows:

$$MSE = \frac{1}{N}\sum_{k=1}^{N}(T_k - Y_k)^2 \tag{16}$$

where $N$ is the number of samples, and $T_k$ and $Y_k$ are the target and its corresponding output (prediction by ANN) respectively at each point $k$.

b)      *ANN Training and Testing*

To adjust the weight and biases - that is, to train the ANN - there are many different algorithms. The algorithm that was used in this work is the Levenberg-Marquardt algorithm [43, 44]. For a vector of weights and biases, $\boldsymbol{Z}_k$ (at iteration $k$), the algorithm iteration to update $\boldsymbol{X}_k$ is:

$$Z_{k+1} = Z_k - [J_k^T J_k + \mu I]^{-1} J_k e_k \qquad (17)$$

where the vector of network error (difference between target and output) is $e$ whose first derivatives with respect to weights and biases are contained in the Jacobian matrix, $J$. $I$ is the identity matrix, and $\mu$ is a positive number called the combination effect. $\mu I$ is added to avoid having a non-invertible matrix $J_k^T J_k$. $\mu$ is called the combination effect because if $\mu I$ is large, this method is called gradient descent method, and if $\mu I$ is small (near zero), it becomes Newton's method. To minimize the error, $\mu$ is initiated with a large value and it gets decreased after each successful iteration. Successful iteration means an iteration that is accompanied by a reduction in MSE.

To test ANN efficiency, the samples are divided into three separate groups: (1)- The training data (most of the data, typically 70% of the whole dataset): these samples are utilized by the network to adjust the weights and biases. (2)- The validation data (typically 15% of the whole data): these samples are used to measure the network generalization (if MSE gets reduced with iterations for training data but it increases for the validation data group, that means the generalization deteriorates). Therefore, the training process stops when generalization stops improving. 3- The testing data (typically 15% of the data): these samples do not influence the training process and they give an independent measure of the ANN predictive capability with a new dataset (one that was not used in the training process).

## 3.2    Random Forest Regression

Random forest regression is composed of a group of regression trees. A regression tree [45] predicts responses to input variables following the decisions in the tree from the root node down to a leaf node (see Fig. 5). Each leaf node contains a possible response, $Y$. Each step in the prediction involves checking the value of one predictor (input). On each predictor, all the possible binary splits are examined. The split that leads to the best result is selected; the optimization criterion is typically minimizing MSE. The split may result in a child node that has very few observations (data); this is why a minimum leaf size (minimum observations per node) is specified *a priori*. For the new child nodes (new leaves), the same procedure is taken to split the new nodes. This process continues again and again. Therefore, the regression tree always has a stopping rule. The stopping rule means that the splitting stops when the number of observations per leaf is equal to the minimum one.

The process of splitting is illustrated in Fig. 5, which represents a simple tree with $x_1$, $x_2$ and $x_3$ as inputs (predictors) with 4 leaves (4 responses). Regression trees are computationally cheap, but they over-fit the data; this is why many trees are used together (tree bagging).

In tree bagging, a random sample of the training data is given to each regression tree so each tree is trained based on a different dataset. Hence, for a new sample (unseen data), each tree will make a different prediction. The prediction of the whole tree bag is the average of all the individual trees' predictions [27].

This method is called the random forest because a random subset of input variables (features) are utilized at each split in each tree (in the tree bag). If the number of the features is $R$, the number of features at each split is typically $R/3$ [46].

The sampling technique used in random forest algorithm is the bootstrap sampling [47]. In this technique, the sample size (for each individual tree) is equal to the original dataset size but the data are chosen with replacement which means that some data points (rows) may be selected more than once and other rows are not chosen at all. In this way, the data given to each tree is different so the trees' predictions will be different and the over-fitting is eliminated when combining trees' results together. On the other side, bootstrapping leads to a sample per tree which looks like the original sample of the original dataset.

The optimum number of trees is determined based on the "Out Of Bag error" (OOB error). OOB error is the average prediction error on a training data point if we used the trees that did not have this data point when bootstrap sampling. It is always desired to achieve minimum OOB error; therefore, the number of trees is chosen based on this criterion. The typical number of trees in random forest algorithms is 100.

## 4. RESULTS AND DISCUSSIONS

This section presents the results of applying the proposed grid-induced error prediction method on the flow inside a lid-driven cubic cavity. We start by validating the high-fidelity fine grid simulations that are utilized to compute the grid-induced error (see Section 4.1). Next, the capability of the proposed CG-CFD approach, to predict the grid-induced error, is assessed given a variation of training and testing data (see Section 4.2). A set of numerical experiments are performed to predict the grid-induced error corresponding to the different velocity components given the scenarios listed in Table 2. Finally, some issues, that need to be addressed to actualize the proposed CG-CFD approach, are discussed in Section 4.3.

## 4.1.    Validation

In this work, simulations are performed for a lid-driven cavity flow with Reynolds numbers ranging from 6,000 to 12,000. Because the grid size requirements increase with increasing Reynolds number, we started by validating OpenFOAM simulation results for a flow whose Reynolds number equals 12,000. Next, the grid, that was used with $Re = 12,000$, is used for the lower Reynolds-number simulations. Fig. 6 compares the velocity profile captured with experiment [48, 49] with *OpenFOAM* simulations. Fig. 6 depicts the velocity profiles computed with different grids: $\Delta = 0.0083$m (corresponding to 120×120×120 grid), $\Delta = 0.0167m$ (corresponding to 60×60×60 grid) and $\Delta = 0.033m$ (corresponding to 30×30×30 grid). It is shown that the computed profile agrees well with the experimental one when adding wall refinement cells. The finest grid ($\Delta = 0.0083$m with wall refinement near the wall) is utilized to gain high-fidelity data for different Reynolds numbers. Using coarse grids, without wall region refinement, results in inaccurate results. In the next section, coarse uniform grids ($\Delta = 0.05$m, $\Delta = 0.033$m, $\Delta = 0.025$m) inaccurate results are corrected with the help of a ML data-driven model.

## 4.2.    Training and Testing Cases

This section goes through the scenarios listed in Table 2 to capture the grid-induced error for the testing flow if the surrogate model is trained with training flows' data. The variable of interest, in the first 9 scenarios, was taken to be the spatial distribution of $U_x$. For the first scenario, both ANN (with 20 neurons) and RFR (with 100 regression tree) were used to predict $\varepsilon_\Delta(U_x)$:

a)    *Scenario I: Reynolds Number Interpolation*

Starting with an ANN, Fig. 7 depicts the reduction in MSE with increasing the number of training iterations (epochs). In each iteration, ANN network weights and biases are re-adjusted to reduce the MSE given the target (actual $\varepsilon_\Delta(U_x)$) and the output (ANN predicted $\varepsilon_\Delta(U_x)$). It is shown that the MSE gets saturated at around 0.0001 after 78 epochs. Ideally, $\text{MSE} = 0$. In Fig. 7, a scatter plot, of the ANN-predicted $\varepsilon_\Delta(U_x)$ vs. the actual $\varepsilon_\Delta(U_x)$, is presented. The training flows, given to ANN, are divided into three groups (training, testing and validating data), as explained in Section 3.1. The dotted diagonal line plot represents a perfect agreement between the targeted and the predicted $\varepsilon_\Delta(U_x)$ while the solid line is the best fit for the available data. The correlation coefficient, $R$ is close to unity which indicates a very good agreement between ANN prediction and target value for all the data groups (training, testing and validating data groups).

While ANN predicts $\varepsilon_\Delta(U_x)$, the variable of interest is $U_x$ (in particular: $U_{x_{f\to\Delta}} = U_{x_\Delta} + \varepsilon_\Delta(U_x)$). A comparison between the actual $U_{x_{f\to\Delta}}$ and the ANN prediction of $U_{x_{f\to\Delta}}$ is illustrated in Fig. 8. The vertical axis refers to the actual $U_{x_{f\to\Delta}}$ ($u_f$ in Fig. 8). The horizontal axis refers to one of three variables: (1) $U_{x_{f\to\Delta}}$ ($u_f$), which corresponds to the solid black line, referring to the perfect case when the ANN-predicted $U_{x_{f\to\Delta}} = U_{x_\Delta} + \varepsilon_\Delta(U_x)$ is exactly equal to the actual one: (2) $U_{x_\Delta}(u_c$ in Fig. 8) which corresponds to the crossed points referring to the coarse grid results compared to fine grid results; (3) ML prediction (circled points), which computes $U_{x_{f \to \Delta}}$ given coarse grid results, $U_{x_\Delta}$ ,and

ANN prediction of $\varepsilon_\Delta(U_x)$. It is shown in Fig. 8 that ML predictions for both training and

testing cases are close to the ideal case. Some circled points deviated from the "ideal"

solid line but most ML predictions (circled points) give visible improvements over the coarse grid results (crossed points). Note that Fig. 8 shows a regression of "big data".

For instance, training flows include data from 3 grids for training: each grid has 27,000

cells with 37 features computed at each cell.

For scenario I, RFR is also utilized for predicting $\varepsilon_\Delta(U_x)$. To optimize the number of trees

in RFR, OOB error (explained in Section 3.2) is minimized. As illustrated in Fig. 9, the

OOB error gets saturated at a minimum value when using 100 regression trees.

As was shown for the ANN, Fig. 10 illustrates the comparison between RFR predictions and coarse grid predictions. RFR training is better than ANN without a single point deviating from the ideal solid line.

b)      *Scenarios II and III: Reynolds Number Extrapolation.*

In the second scenario, the testing flow case gets more challenging because the testing

case is outside the training flows' range. Prediction of the grid-induced error using ANN

and RFR are presented in Fig. 11 and Fig. 12. ML capability of prediction in scenario II is not as good as the scenario I. However, ML predictions are still better than coarse grid results. RFR predictions do not deviate much away from the ideal line.

In Table 3, a comparison between ANN and RFR efficiency, in different scenarios, is presented. The comparison is presented in terms of ML error, $E_{ML}$, defined by Eq. (6), and the computational cost. In both scenarios (I and II), RFR leads to smaller mean and maximum $E_{ML}$ compared to ANN. RFR is also computationally cheaper than ANN. Thus, for the next scenarios, only RFR is utilized. In scenario III (extrapolation to a lower Reynolds number, plotted in Fig. 13), RFR still results in good predictions.

c)      *Scenarios IV, V and VI: Grid Size Interpolation and Extrapolation.*

The focus of scenarios IV, V and VI is the grid size. RFR predictions for these scenarios are illustrated in Fig. 14, Fig. 15, and Fig. 16. RFR performs very good data training for the three scenarios. RFR regression for the testing case is not as good as the training flows' data. RFR predictions for the testing cases (interpolation or extrapolation to a finer or a coarser grid) are still close to the ideal solid line. In Table 3, the mean $E_{ML}$ is around 0.4 for the three scenarios while the maximum $E_{ML}$ increases when extrapolation to a finer grid ($E_{ML} = 4$) and decreases in extrapolation to a coarser grid ($E_{ML} = 2$).

d)      *Scenarios VII, VIII and IX: Reynolds Number and Grid Size Interpolation and Extrapolation.*

Scenarios VII, VIII and IX are the most challenging ones because the testing flows differ from the training flows in <u>both</u> Reynolds number and grid size. As illustrated in Fig. 17, Fig. 18 and Fig. 19, most RFR predictions are very close the ideal solid line. According to Table 3, for the scenarios VII, VIII and IX, the mean $E_{ML}$ is 0.4 or lower, and the maximum $E_{ML}$ is not high compared to the range of $E_{ML}$ for previous scenarios which are less demanding.

*e)        Other Variables $U_y$ and $U_z$*

All the previous numerical experiments were performed assuming that the variable of interest is $U_x$ (which corresponds to the velocity direction of the cavity lid motion). In this section, the variables $U_y$ and $U_z$ are considered with scenarios VII and VIII (in Table 2). Both scenarios VII and VIII were selected as they represent the more general scenarios (with testing flows' grid size and Reynolds number different from the training flows). Note that the feature vector defined in Eq. (4) remains unchanged when the output of interest ($\varepsilon_\Delta(U_x)$) is replaced by $\varepsilon_\Delta(U_y)$ and $\varepsilon_\Delta(U_z)$.

In Fig. 20 and Fig. 21, RFR predictions, for the velocity $U_y$, are compared to the true values. In both cases, RFR prediction gives an improvement over the coarse grid predictions for most points. On the other hand, RFR performance is poor when predicting $U_z$ in Fig. 22 and Fig. 23 . RFR predictions for $U_z$ give no improvement over the coarse grid results.

The performance of RFR in predicting $U_y$ and $U_z$ is assessed in Table 3 in terms of $E_{ML}$. For the velocity $U_y$, $E_{ML}$ is comparable to (a little bit higher than) $E_{ML}$ when computing $U_x$. On the other hand, for the velocity $U_z$, ML errors for the testing data are very high compared to $U_x$ and $U_y$. For $U_z$, the correlation between the fine grid velocity and the coarse grid velocity is very poor. $U_z$ is very small (and less significant) compared to the other velocity components, so it is sensitive to any ML error.

*f)*     *Data Convergence*

In the proposed approach, the flow of interest (testing flow) simulation grid-induced error is predicted given a surrogate model based on data from the available training flows. It is necessary to explore how this surrogate model accuracy will be impacted by adding or removing some data. The surrogate model should give better performance with more training flows. This is called "data convergence".

Four studies of data convergence are presented in Table 4 (assuming that $U_x$ is the variable of interest): In the first study, all the training and testing flows have the same grid size ($\Delta= 1/30$ ) while the Reynolds number changes. Adding more training data with Reynolds number closer to the testing data leads to a lower $E_{ML}$. In the second study, adding more training data with Reynolds number far from testing data, led to lower mean $E_{ML}$ and little change in the maximum $E_{ML}$. The studies III and IV focus on adding data with different grid size while the Reynolds number is fixed. The studies III and IV led to results similar to the first and the second studies.

## 4.3.    Open Issues

The case study results, presented in this paper, indicate potential of using machine learning algorithms for prediction of the CG- CFD error. The study also points to a range of challenges and open issues that need to be addressed before the proposed method could become practical:

- Surrogate model validation range: The validation range of the surrogate model, developed by ML, is yet to be established. It is expected that the capability of the surrogate model is related to the "closeness" between the training flows and the testing flows. A metric for this closeness may serve as *a priori* assessment for the surrogate model prediction confidence.

- Flow complexity: Compared to flows in real world applications, the studied case (lid-driven cavity flow) is three-dimensional and turbulent while isothermal. Non-isothermal flow needs to be studied to assess the capability of the proposed method. This would require additional features (e.g. temperature gradients) plus potentially more sophisticated ML algorithms (e.g., deep learning [50]). Additionally, the flow studied here is quasi steady state while the flows in many engineering applications may be transient. However, in the proposed approach, a quasi-steady-state approach is adopted (the flow variables are corrected locally, based on local flow features), for the local time scale can be assumed to be much shorter than characteristic time scale of transient in engineering systems of interest. More generally, time derivatives terms may be involved as flow features to account for the transient behavior.

- Conservation and Galilean invariance requirements: In the case study, velocity components computed by CG-CFD are corrected separately, without enforcing conservation and Galilean invariance. Consequently, ML predictions are dependent on the orientation of coordinate frame. This limits the predictive capability especially if the training flows and testing flows have different geometries.

## 5.    CONCLUSIONS

High-resolution results from simulations or experiments produce an enormous amount of data (spatio-temporal distribution of field variables of interest; typical grid sizes may exceed billions of data points over millions of time steps). These "big data" are not optimally usable because, for each new scenario of interest, a sufficiently fine grid CFD simulation needs to be performed. This traditional approach is computationally overwhelming for simulation of thermal-fluid processes in many practical large-scale and engineering applications.

In the present work, CG-CFD simulations are performed and the CG-CFD induced error is learned by a surrogate model constructed by ML algorithms. The surrogate model is trained given the available fine grid and coarse grid data. Both fine and coarse grid data were performed with the same set of conservation equations (no turbulence modeling). Hence, it was assumed that coarsening the grid leads to higher discretization error while the model discrepancy is negligible. The coarse grid-induced local error distribution was predicted, given features computed from the coarse grid simulations. The function that relates the error to the features is constructed using ML techniques: ANN and RFR. The proposed method was applied to analyze three-dimensional flow inside a lid driven cavity.

The proposed approach was found to be capable of correcting the coarse grid results, obtaining reasonable predictions for new cases (having different Reynolds number and/or being computed using different grid sizes). The velocity components of interest in the new cases, were corrected over the whole solution domain. The surrogate model performance improved when adding data that are more relevant and it is generally insensitive when less-relevant data are incorporated into the training.

While sufficiently accurate predictions were obtained in both algorithms using ANN and RFR, the one using the RFR technique results in predictions that are more accurate and with higher computational efficiency. The CG-CFD method performance is promising, and has the potential to enable more computationally affordable simulations, which benefit from the available and growing high-fidelity simulation and experimental data. The method still needs further assessment in scenarios when the testing and the training flow have different geometries or different physics.

**ACKNOWLEDGMENT**

This work is supported through the INL Laboratory Directed Research & Development (LDRD) program under DOE Idaho Operations Office Contract DE-AC07-05ID14517.

**ABBREVIATIONS**

| | |
|---|---|
| ALES | Autonomic Sub-grid Large Eddy Simulation |
| ANN | Artificial Neural Network |
| CFD | Computational Fluid Dynamics |
| CG | Coarse Grid |
| CG-CFD | Coarse-Grid Computational Fluid Dynamics |
| DNS | Direct Numerical Simulation |
| HPC | High Performance Computing |
| LES | Large Eddy Simulation |
| MATLAB | Matrix Laboratory |
| ML | Machine Learning |
| MSE | Mean Squared Error |
| NS | Navier Stokes |
| OOB | Out Of Bag |
| PIML | Physics Informed Machine Learning |
| PISO | Pressure Implicit Splitting of Operators |
| RANS | Reynolds Averaged Navier Stokes |
| RFR | Random Forest Regression |
| SGS | Sub-Grid Scale |
| TKE | Turbulence Kinetic Energy |

**NOMENCLATURE**

| | |
|---|---|
| $E_{ML}$ | Machine learning error metric |
| $F$ | Surrogate function |
| $H$ | cavity characteristic length (height) |
| $R$ | Correlation coefficient |
| $Re$ | Reynolds number. |
| $Re_\Delta$ | Computational grid cell based Reynolds number |
| $U$ | Velocity vector. |
| $U_x, U_y, U_z$ | Velocity components in the Cartesian coordinates. |
| $V$ | Volume of computational cell. |
| $X$ | Features (inputs) in machine learning algorithm |
| $\tilde{X}$ | The normalized features |
| $p$ | Kinematic pressure. |
| $t$ | Time |
| $x, y, z$ | Cartesian coordinates. |

Greek letters

| | |
|---|---|
| $\Delta$ | Grid spacing or cell length. |
| $\psi$ | A number or a set of numbers which characterize the flow pattern |
| $\beta$ | A spatially distributed multiplicative discrepancy term. |
| $\varepsilon_\Delta(\varphi)$ | Grid-induced error when computing a variable $\varphi$ |
| $\tilde{\varepsilon}_\Delta$ | The normalized grid induced error |
| $\nu$ | Kinematic viscosity. |
| $\sigma$ | Standard deviation |
| $\varphi$ | Flow variable. |
| $\varphi_f$ | Variable computed by fine grid. |
| $\varphi_f^{tr}$ | Variable computed by fine grid for the training flows |

$\varphi_\Delta$          Variable computed by a grid, $\Delta$.

$\varphi_\Delta^{tr}$          Variable computed by a grid, $\Delta$ for the training flows

$\varphi_\Delta^{te}$          Variable computed by a grid, $\Delta$ for the testing flows

$\varphi_{f \to \Delta}$          High-fidelity result mapped on a coarse-grid $\Delta$

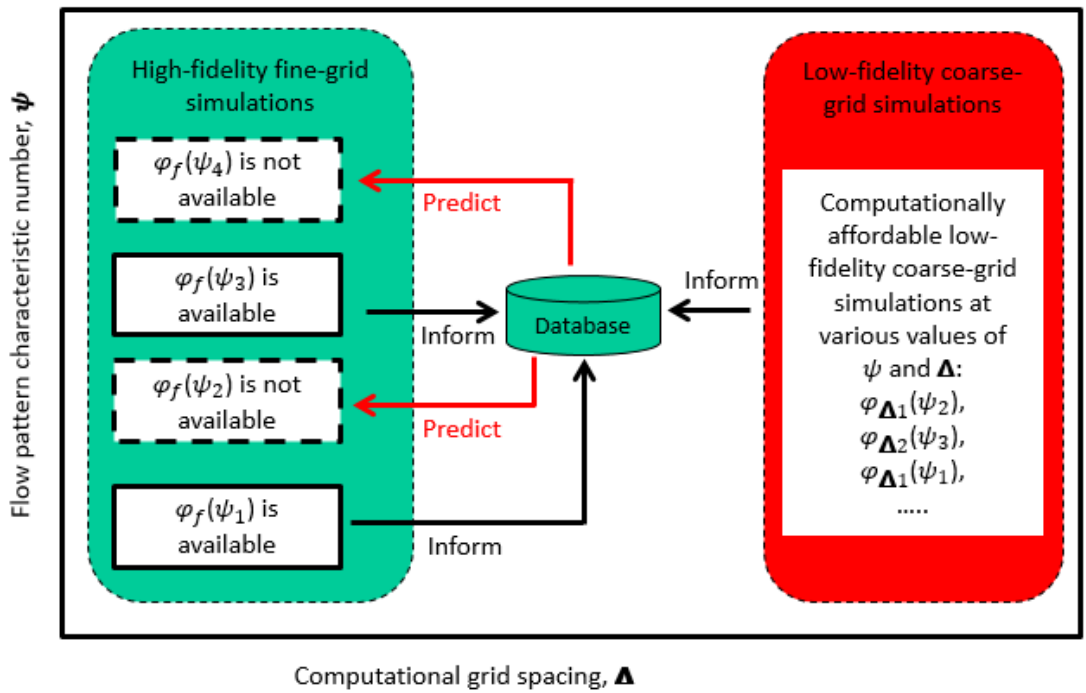

Fig. 1. Work-flow for data generation in the CG-CFD problem.

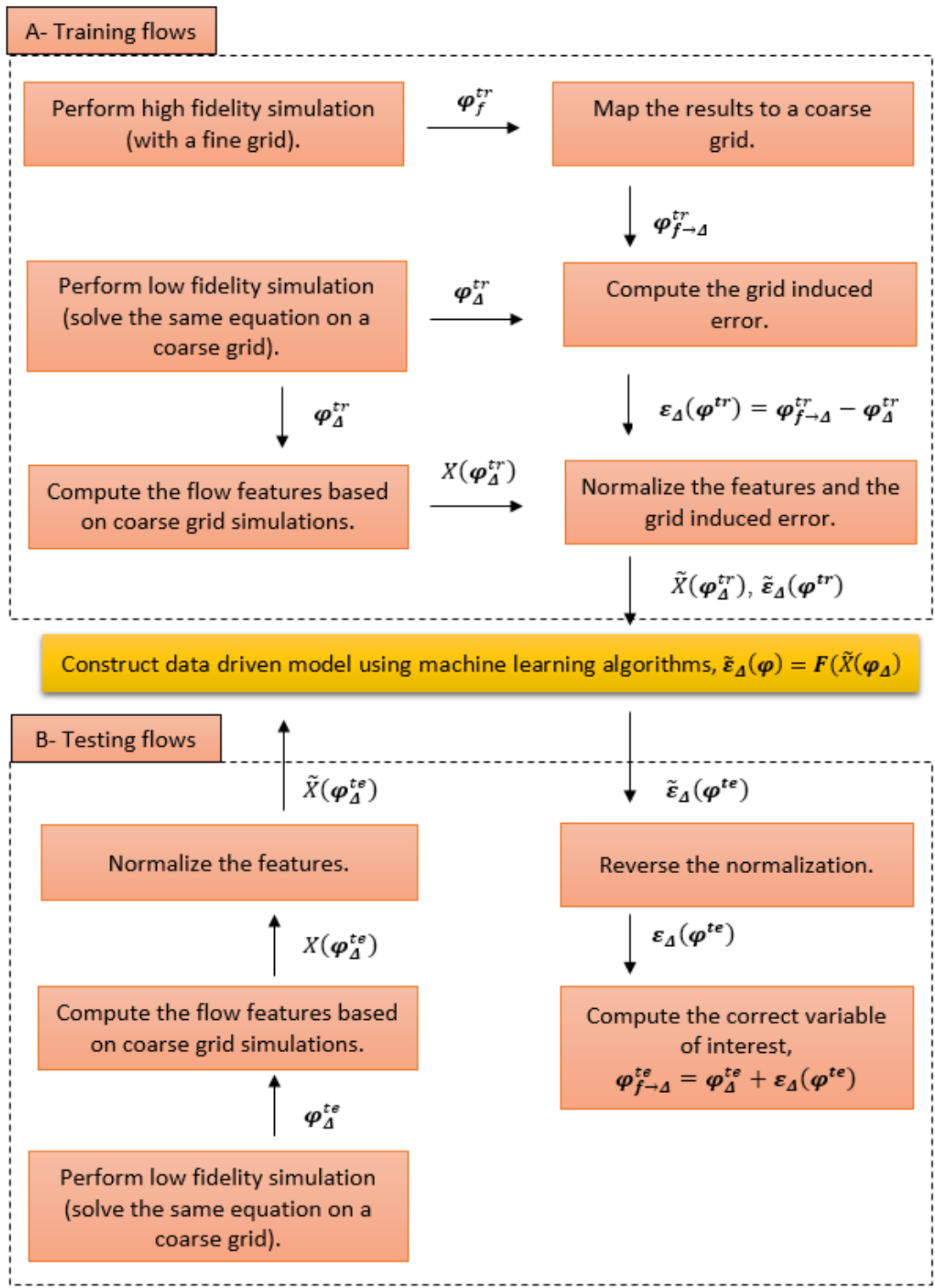

Fig. 2. A method to predict CG-CFD error using ML algorithms.

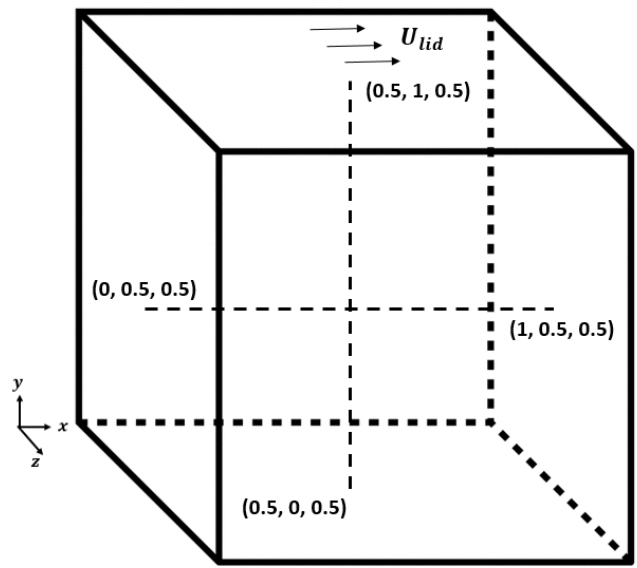

Fig. 3. Lid driven cubic cavity. The lid velocity is parallel to $x$ axis.

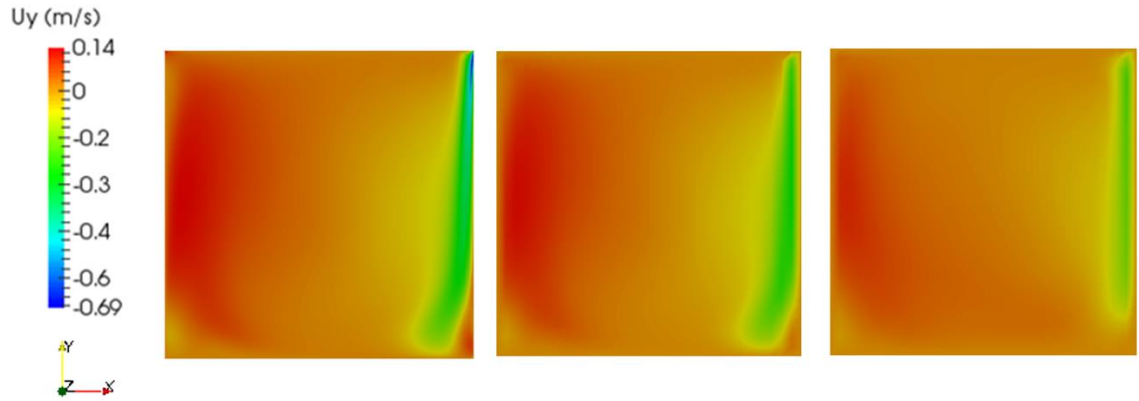

Fig. 4. A snapshot of the velocity ($U_y$) profile in a three-dimensional quasi steady state turbulent flow in a lid driven cavity. The lid is moving in $x$ direction. The profile is computed with a fine grid (left) and a coarse-grid (right). The fine grid result is mapped to a coarse-grid (in the middle).

Table 1. Normalization factors for different variables.

| Feature | Normalization factor |
|---|---|
| $\Delta$ | $H$ |
| $\left(\dfrac{du_i}{dx_j}\right)$ | $H/U_{lid}$ |
| $\left(\dfrac{d^2 u_i}{dx_j{}^2}\right)$ | $H^2/U_{lid}$ |
| $\varepsilon_\Delta(U_x), \varepsilon_\Delta(U_y), \varepsilon_\Delta(U_z)$ | $U_{lid}$ |

Table 2. Training and testing flows. $Re$ and $\Delta$ are the Reynolds number and the grid spacing, respectively.

| Scenario | Training flows | | Testing flow | |
|---|---|---|---|---|
| | $Re$ | $\Delta\,(m)$ | $Re$ | $\Delta\,(m)$ |
| I: $Re$ interpolation. | $\{6000, 8000, 12000\}$ | $\dfrac{1}{30}$ | $10000$ | $\dfrac{1}{30}$ |
| II: $Re$ extrapolation (higher $Re$). | $\{6000, 8000, 10000\}$ | $\dfrac{1}{30}$ | $12000$ | $\dfrac{1}{30}$ |
| III: $Re$ extrapolation (lower $Re$). | $\{8000, 10000, 12000\}$ | $\dfrac{1}{30}$ | $6000$ | $\dfrac{1}{30}$ |
| IV: $\Delta$ interpolation. | $12000$ | $\left\{\dfrac{1}{40}, \dfrac{1}{20}\right\}$ | $12000$ | $\dfrac{1}{30}$ |
| V: $\Delta$ extrapolation (finer grid). | $12000$ | $\left\{\dfrac{1}{30}, \dfrac{1}{20}\right\}$ | $12000$ | $\dfrac{1}{40}$ |
| VI: $\Delta$ extrapolation (coarser grid). | $12000$ | $\left\{\dfrac{1}{40}, \dfrac{1}{30}\right\}$ | $12000$ | $\dfrac{1}{20}$ |
| VII: $Re$ and $\Delta$ interpolation. | $\{8000, 12000\}$ | $\left\{\dfrac{1}{40}, \dfrac{1}{20}\right\}$ | $10000$ | $\dfrac{1}{30}$ |
| VIII: $Re$ and $\Delta$ extrapolation (higher $Re$ and finer grid) | $\{8000, 10000\}$ | $\left\{\dfrac{1}{30}, \dfrac{1}{20}\right\}$ | $12000$ | $\dfrac{1}{40}$ |
| IX: $Re$ and $\Delta$ extrapolation (lower and coarser grid). | $\{10000, 12000\}$ | $\left\{\dfrac{1}{40}, \dfrac{1}{30}\right\}$ | $8000$ | $\dfrac{1}{20}$ |

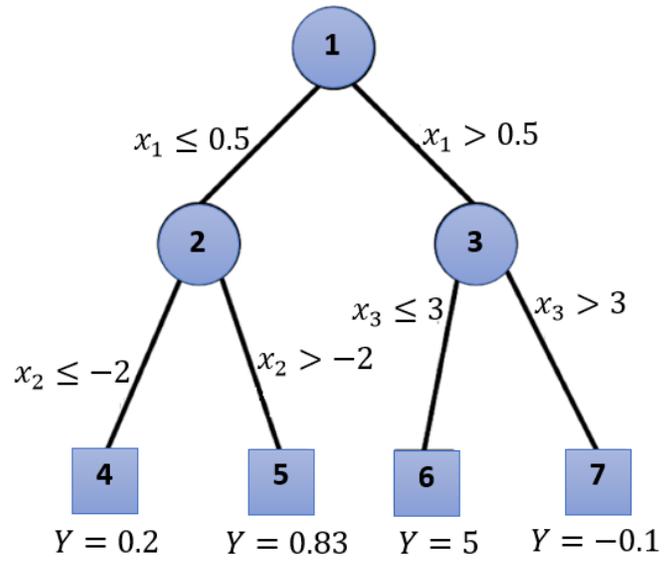

Fig. 5. Regression tree.

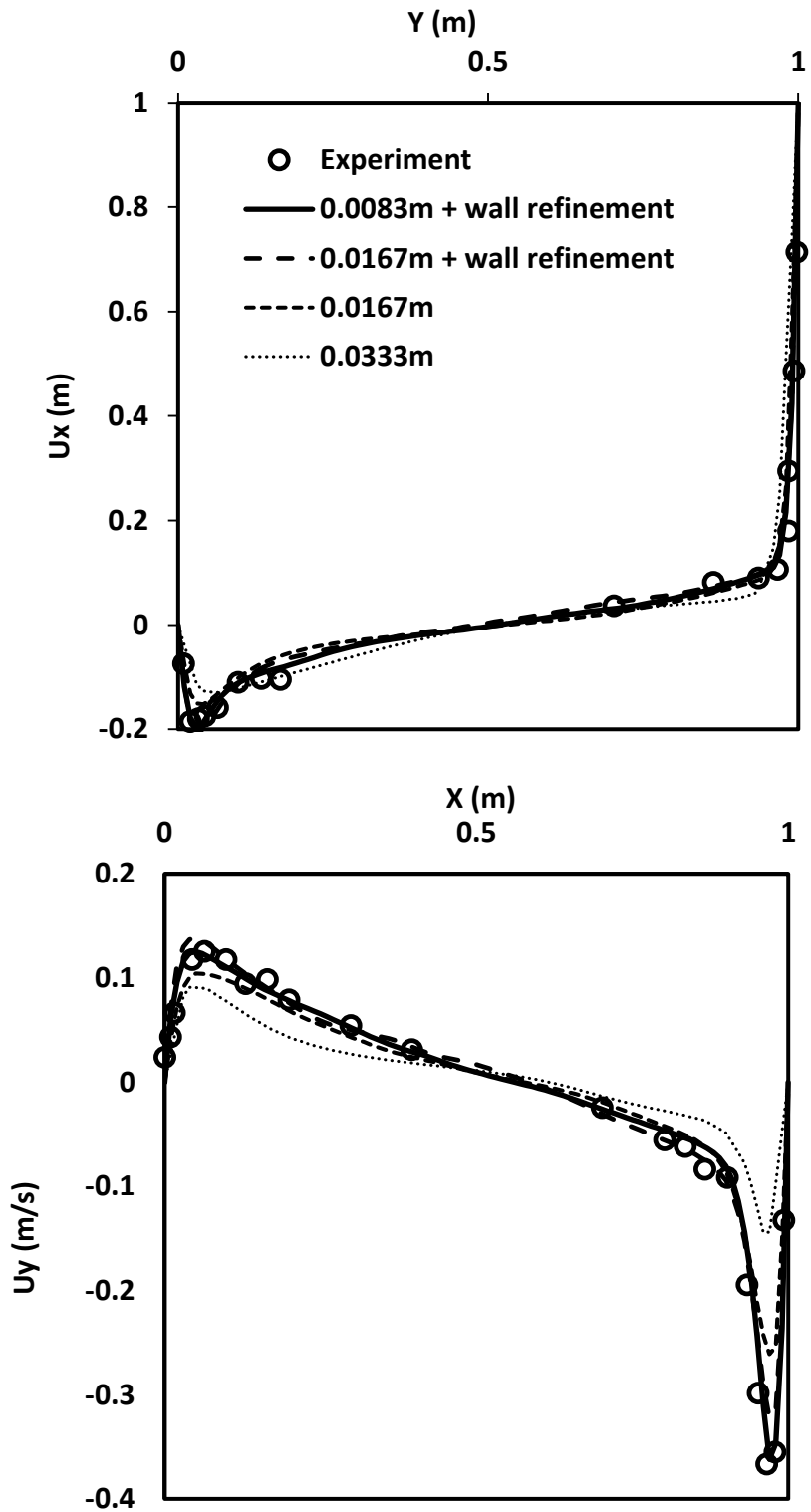

Fig. 6. The axial profile ($y$ direction) for velocity, $U_x$ (top) and the axial profile ($x$ direction for the velocity, $U_y$ (bottom) at $Re = 12000$.

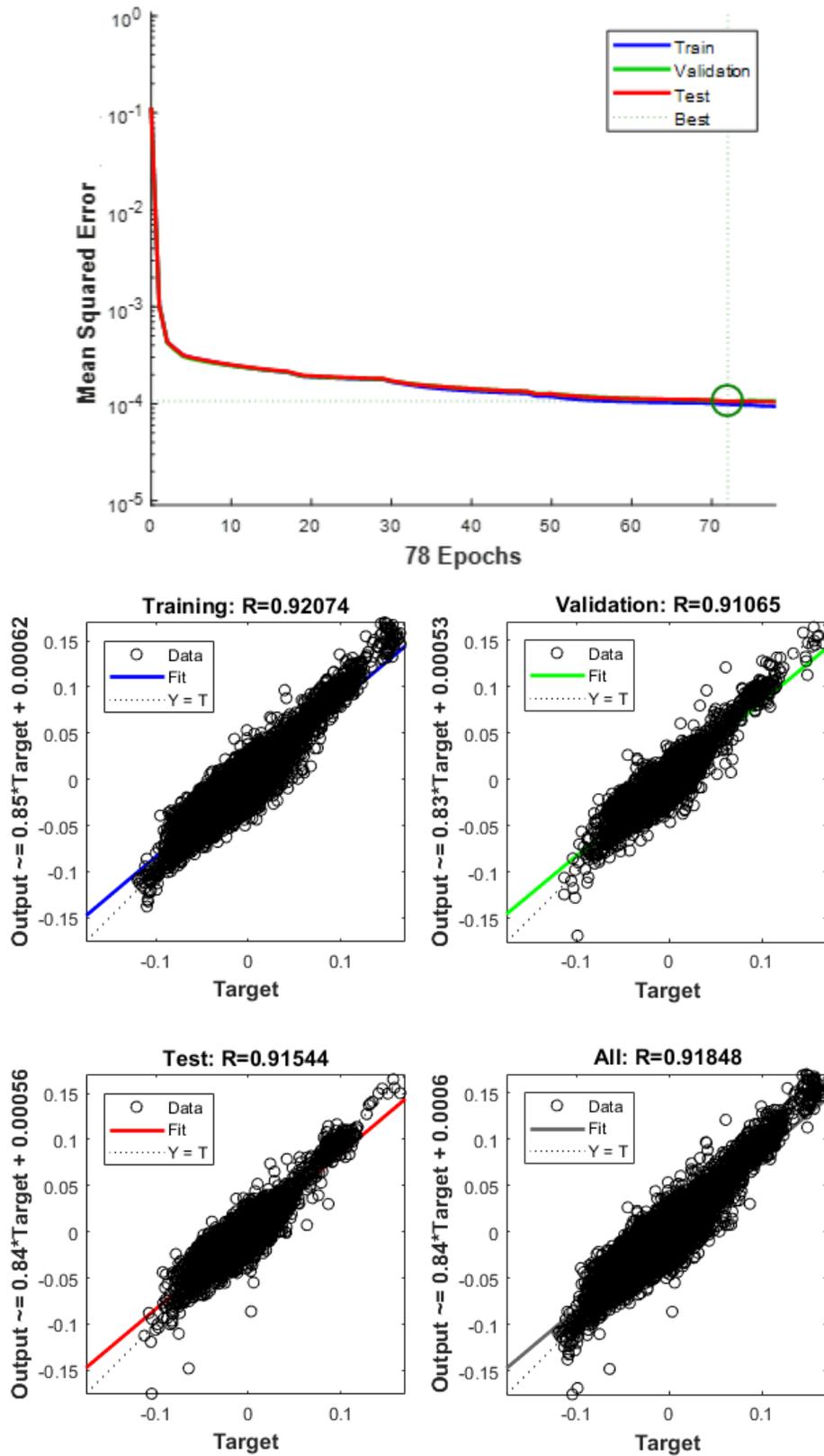

Fig. 7. ANN performance in scenario I (training flows). MSE vs. the number of iterations (epochs) (above). $\varepsilon_\Delta(U_x)$ expected by ANN vs. the actual $\varepsilon_\Delta(U_x)$ (below).

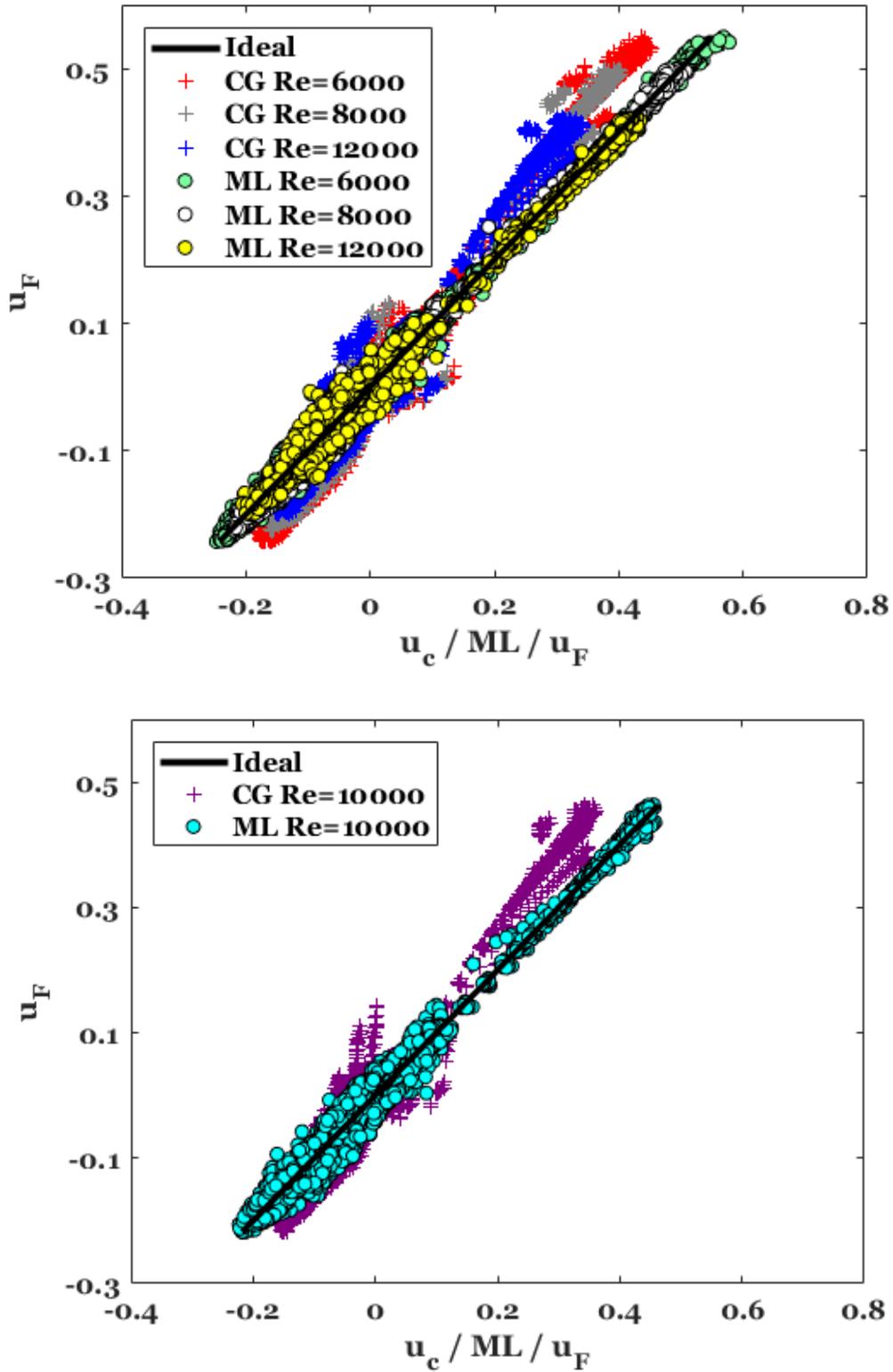

Fig. 8. Scenario I for $U_x$ (by ANN). $\Delta = 0.033 m$. Training data (above) and testing data (below). CG: Coarse grid predictions. ML: Machine learning predictions.

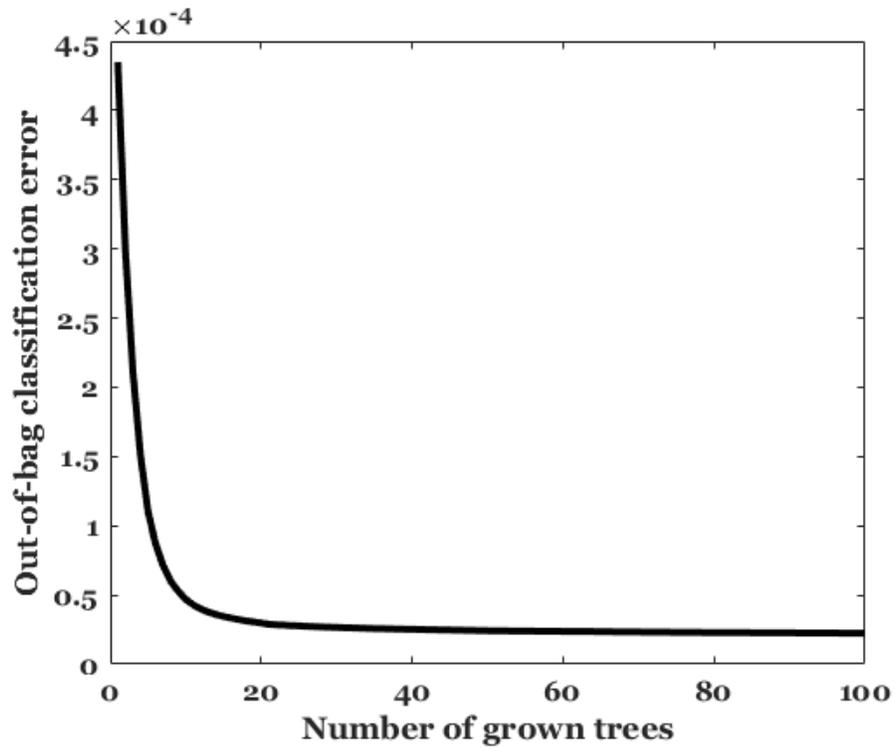

Fig. 9. RFR OOB error decreases with increasing number of trees.

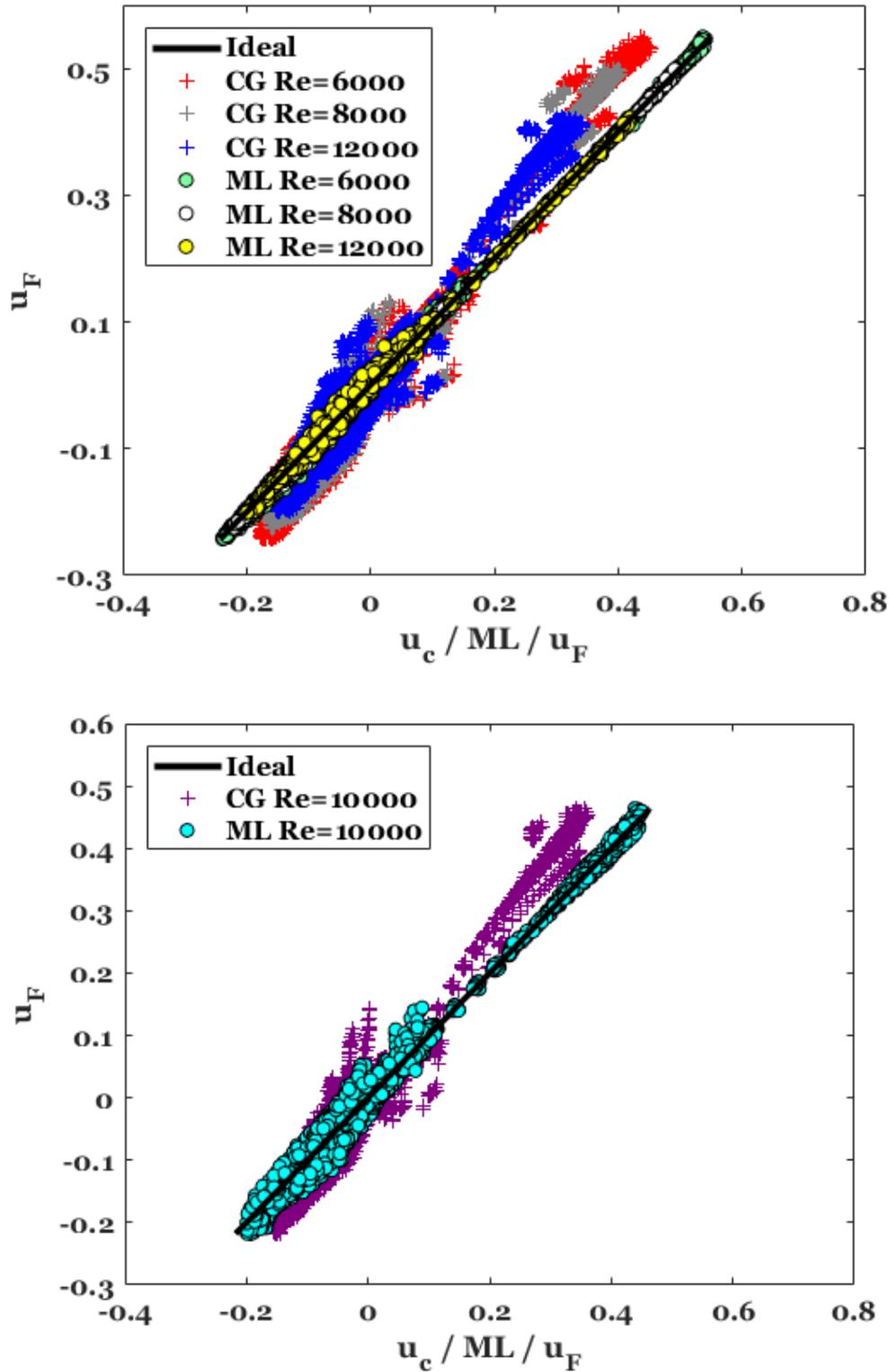

Fig. 10. Scenario I for $U_x$ (by RFR). $\Delta = 0.033m$. Training data (above) and testing data (below).

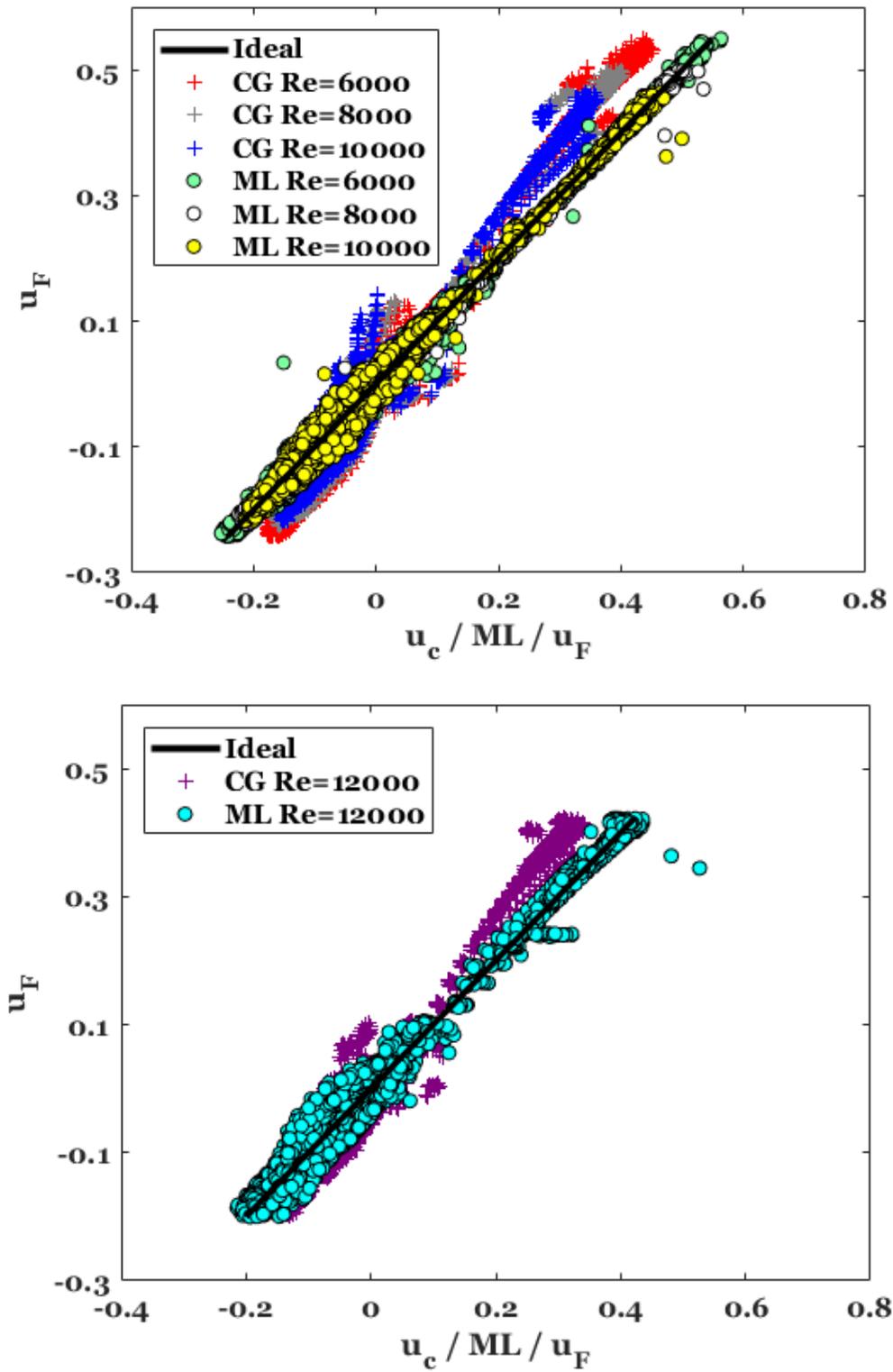

Fig. 11. Scenario II for $U_x$ (by ANN). $\Delta = 0.033m$. Training data (above) and testing data (below).

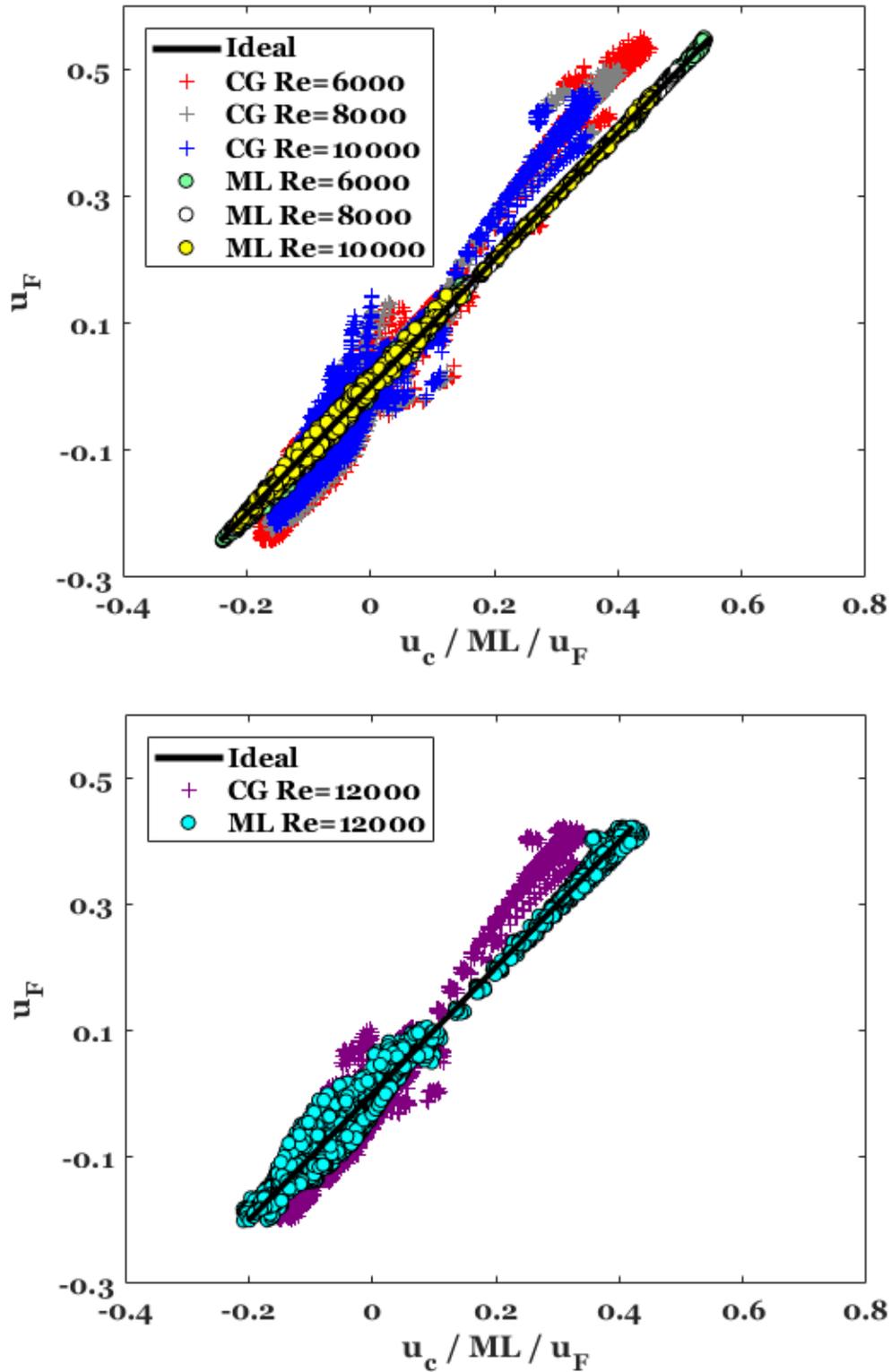

Fig. 12. Scenario II for $U_x$ (by RFR). $\Delta = 0.033m$. Training data (above) and testing data (below).

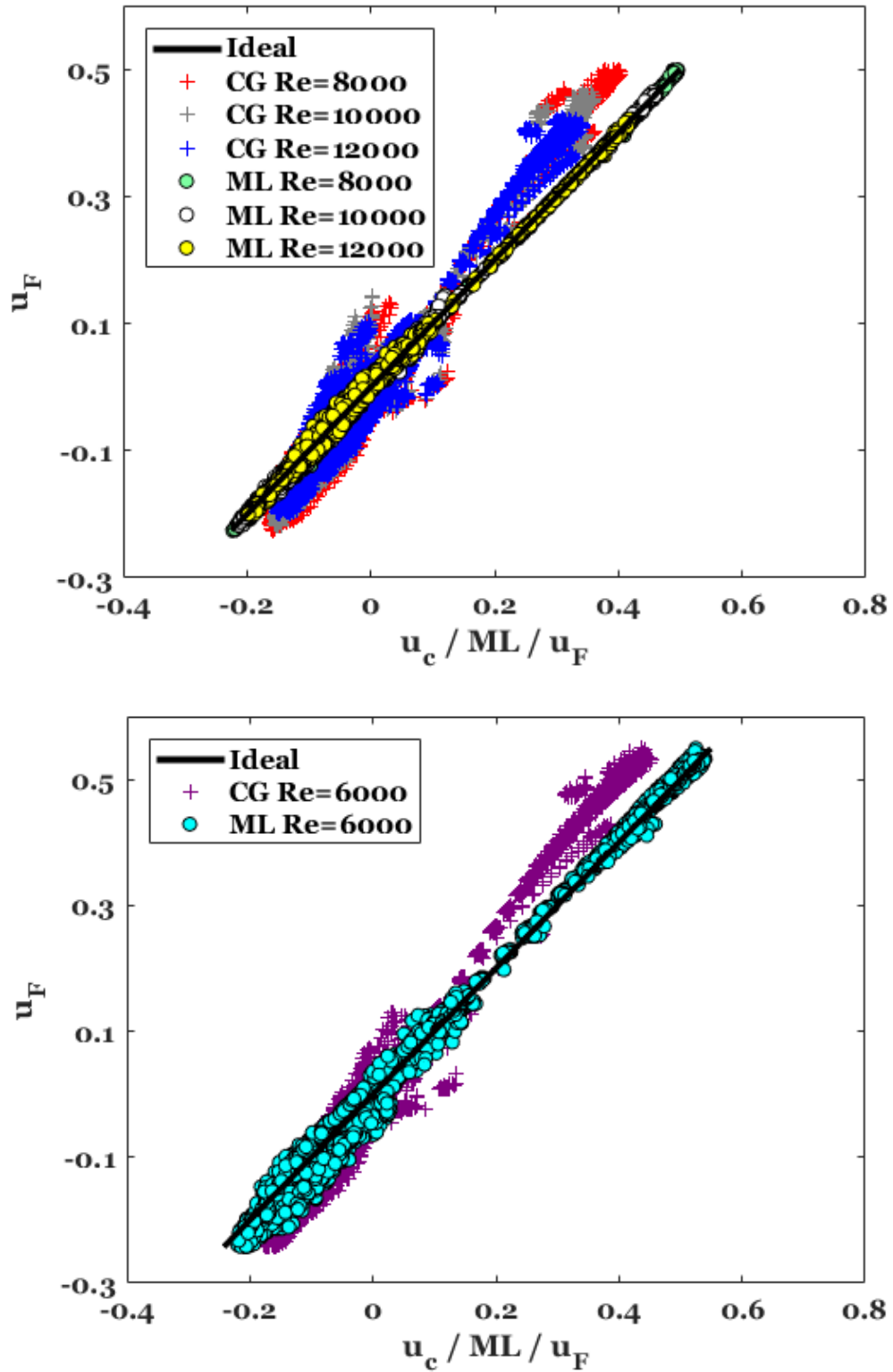

Fig. 13. Scenario III for $U_x$ (by RFR). $\Delta = 0.033m$. Training data (above) and testing data (below).

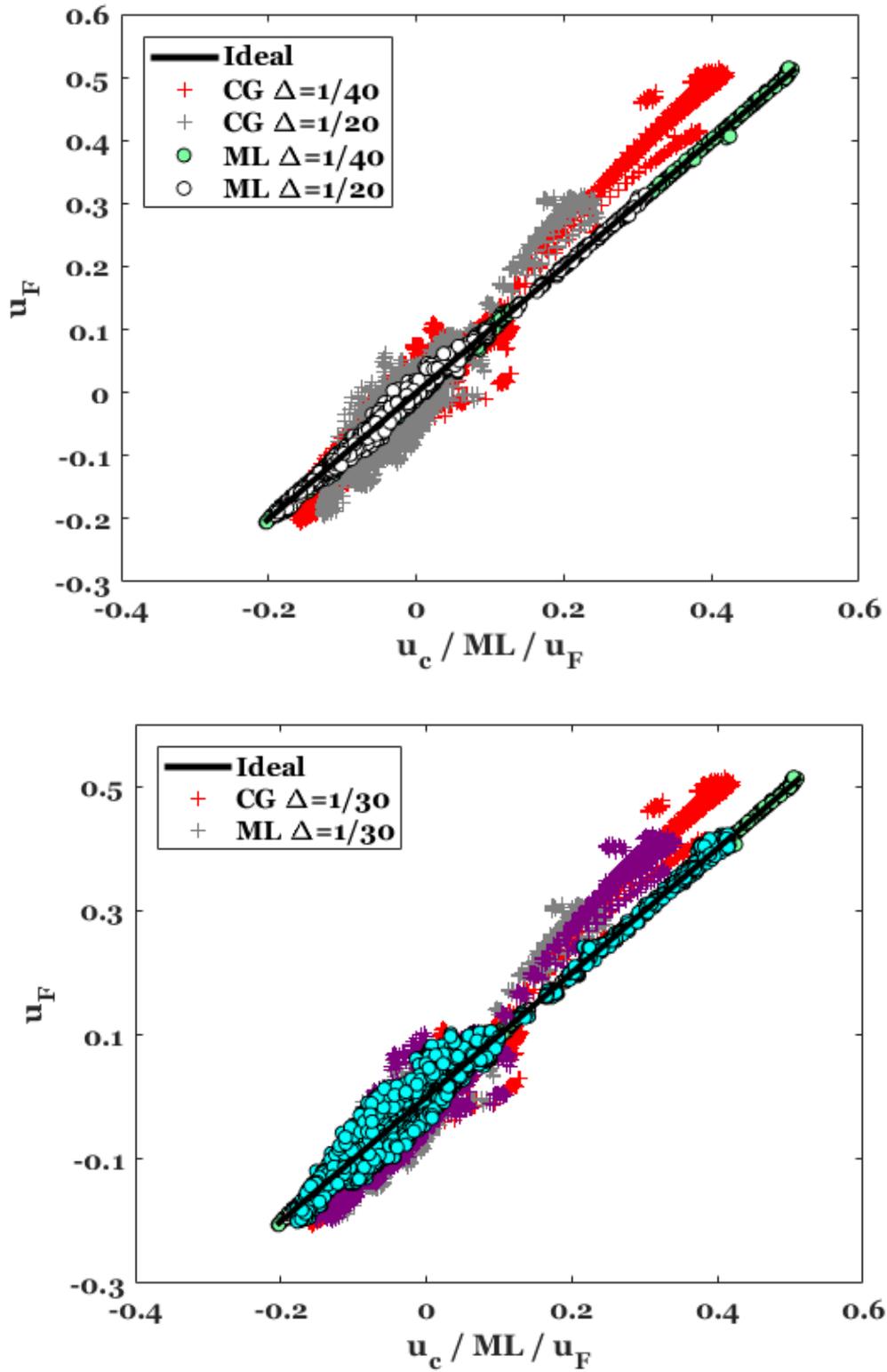

Fig. 14. Scenario IV for $U_x$ (by RFR). $Re = 12000$. Training data (above) and testing data (below).

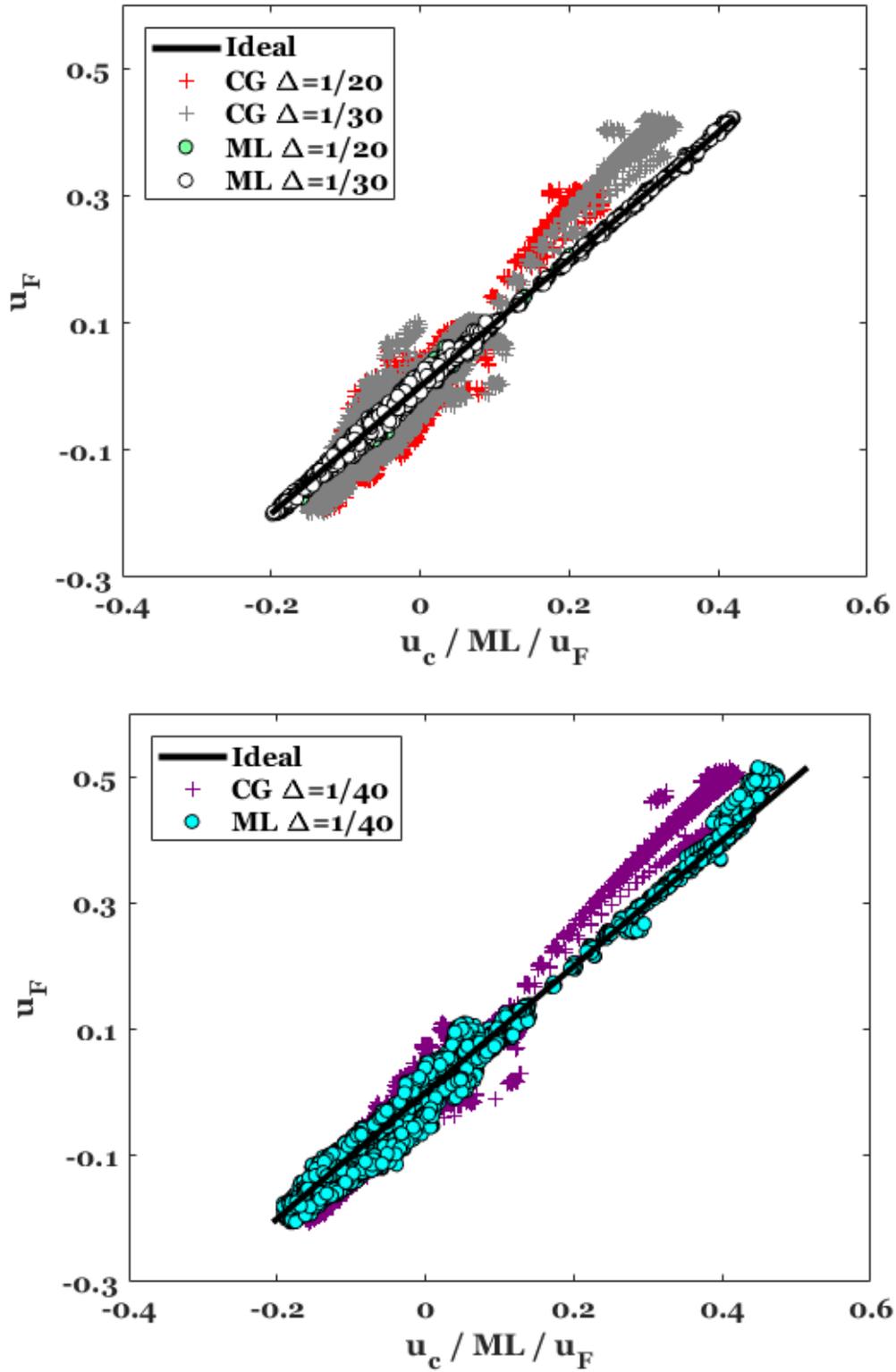

Fig. 15. Scenario V for $U_x$ (by RFR). $Re = 12000$. Training data (above) and testing data (below).

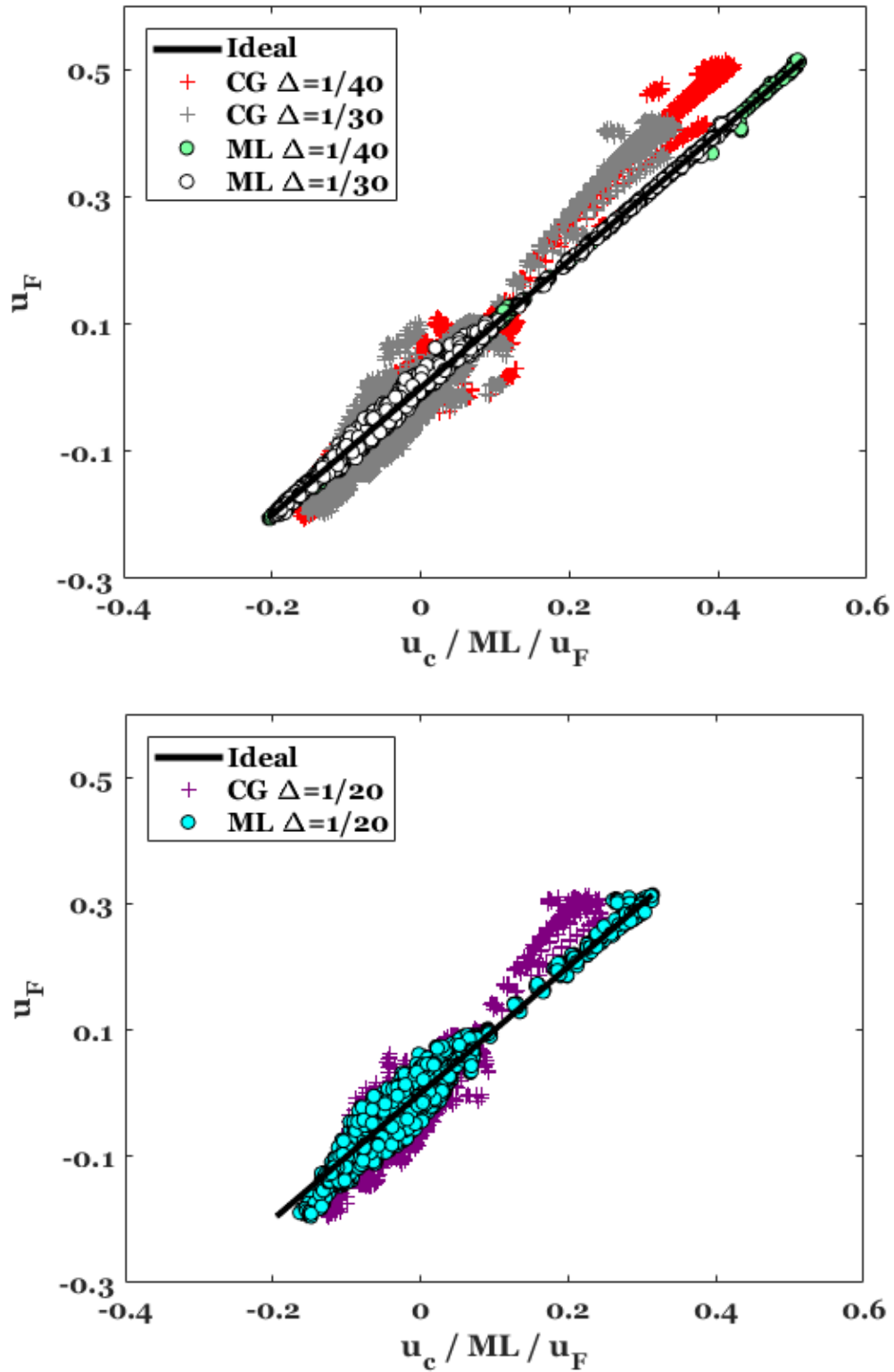

Fig. 16. Scenario VI for $U_x$ (by RFR). $Re = 12000$. Training data (above) and testing data (below).

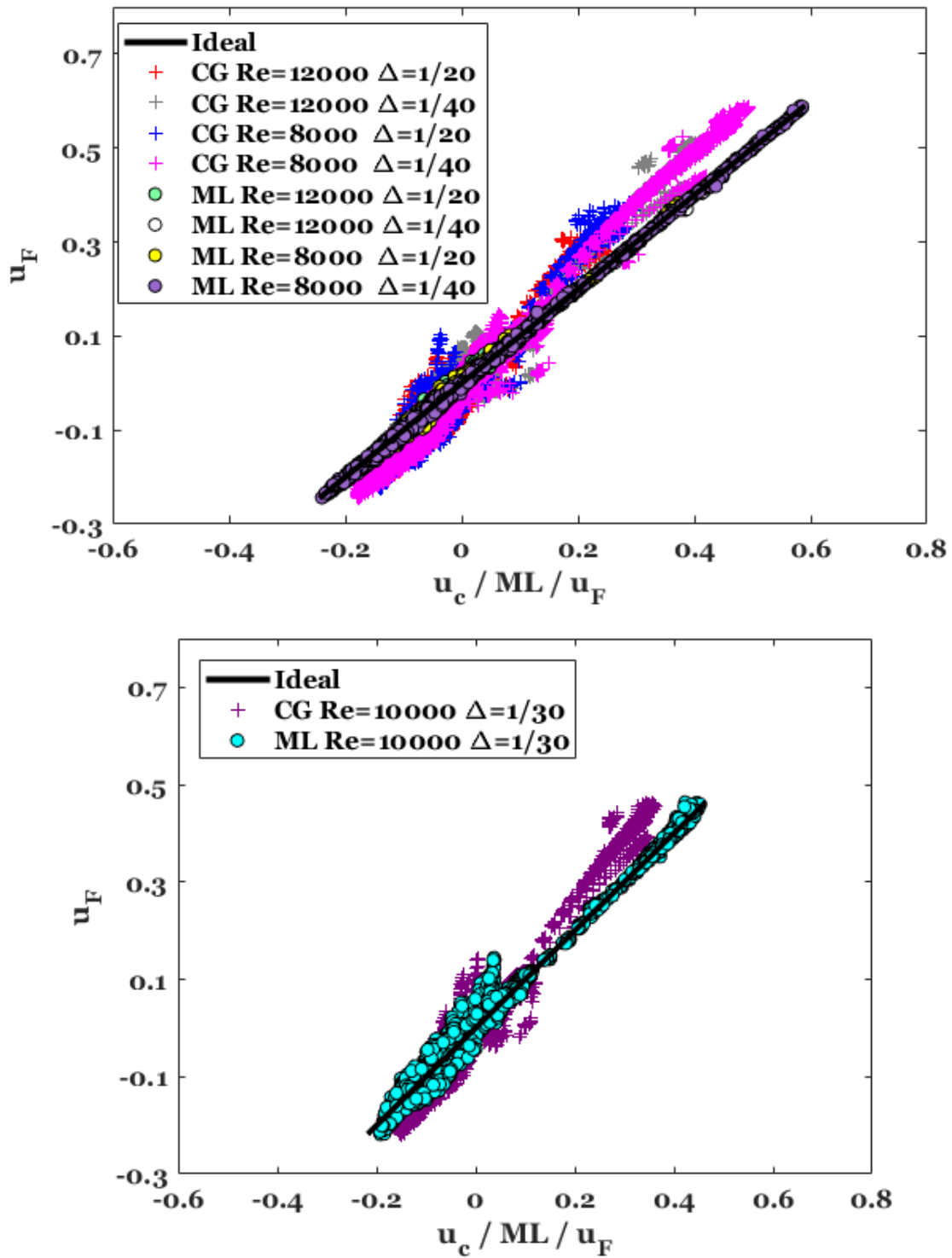

Fig. 17. Scenario VII for $U_x$ (by RFR). Training data (above) and testing data (below).

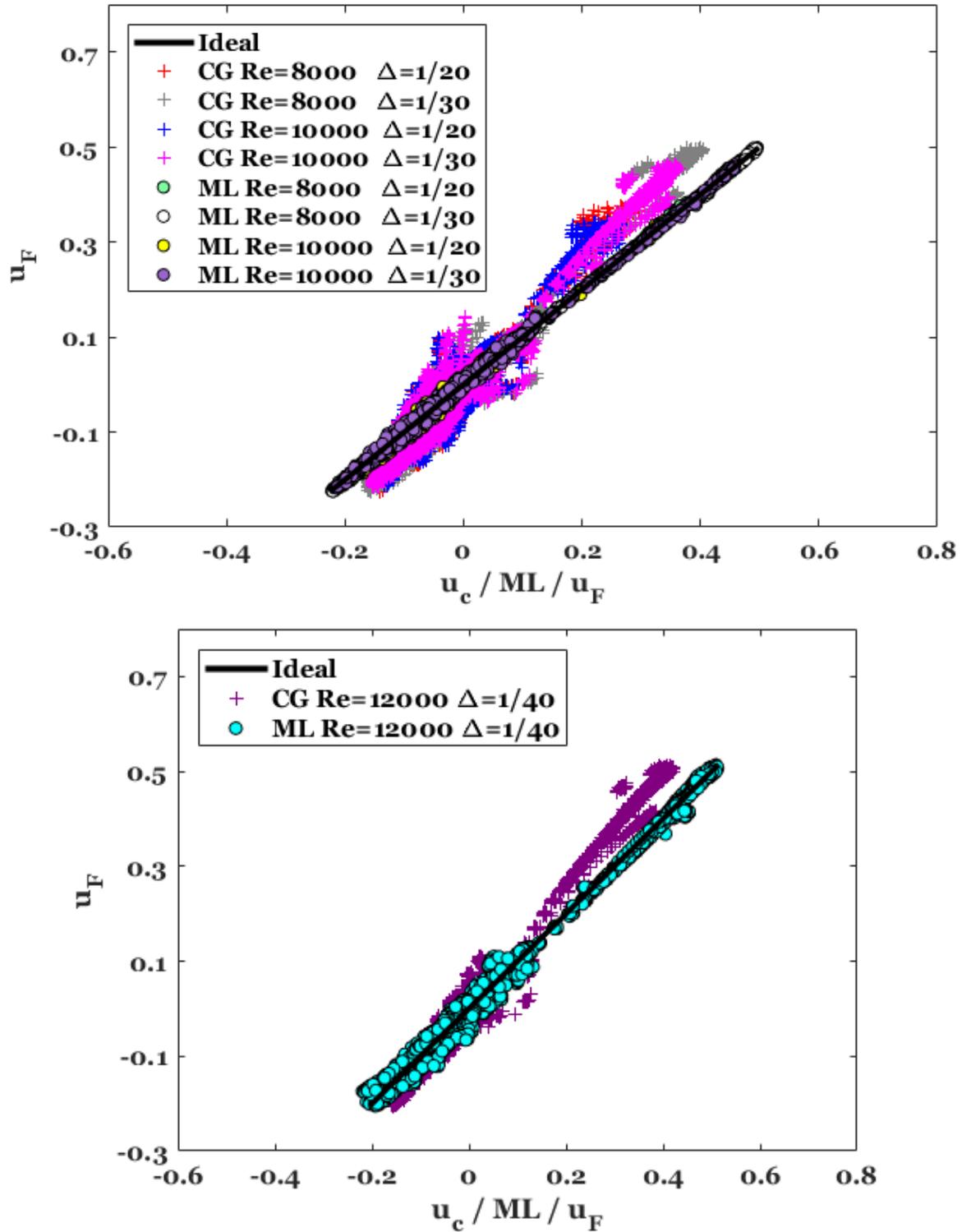

Fig. 18. Scenario VIII for $U_x$ (by RFR). Training data (above) and testing data (below).

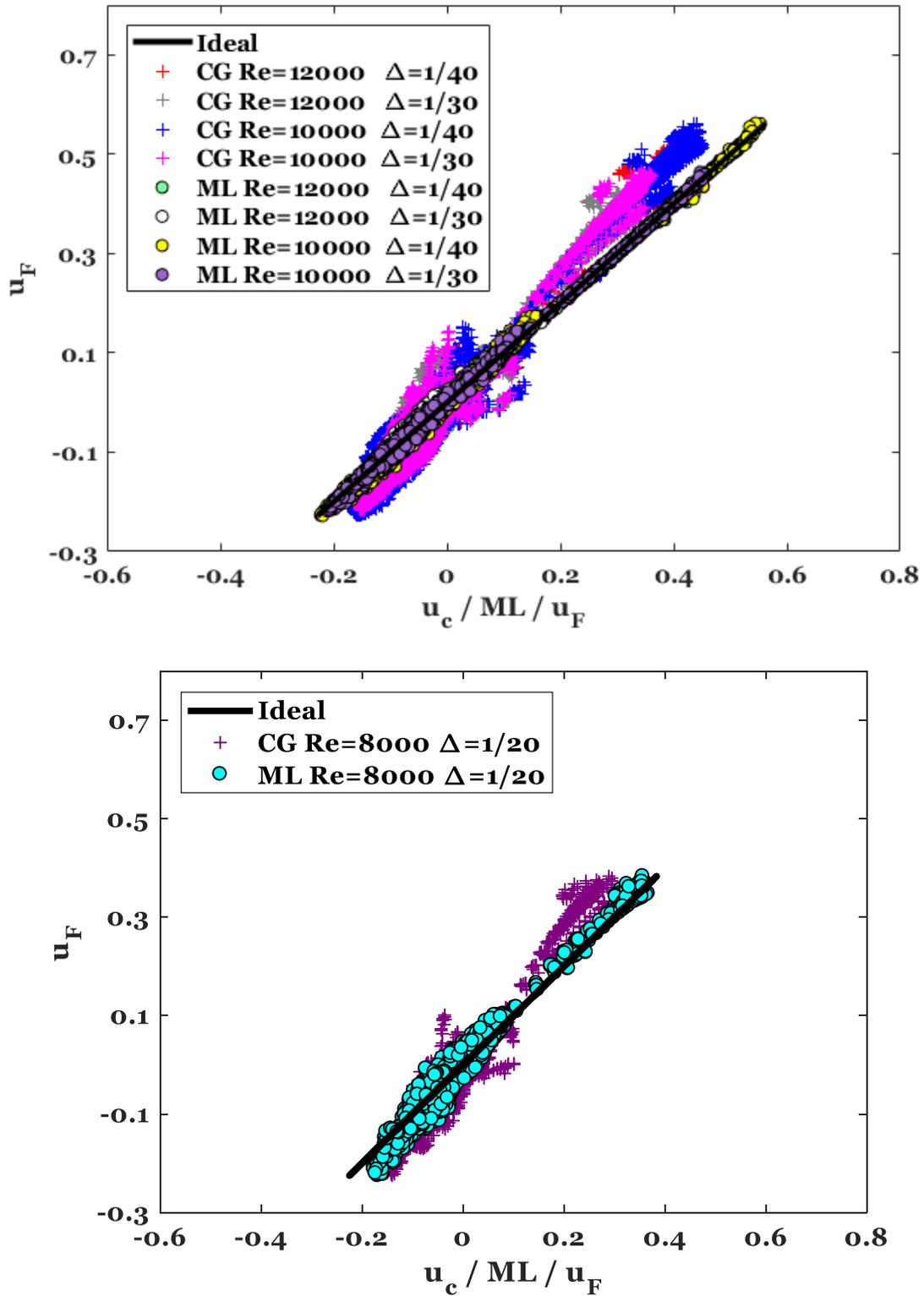

Fig. 19. Scenario IX for $U_x$ (by RFR). Training data (above) and testing data (below).

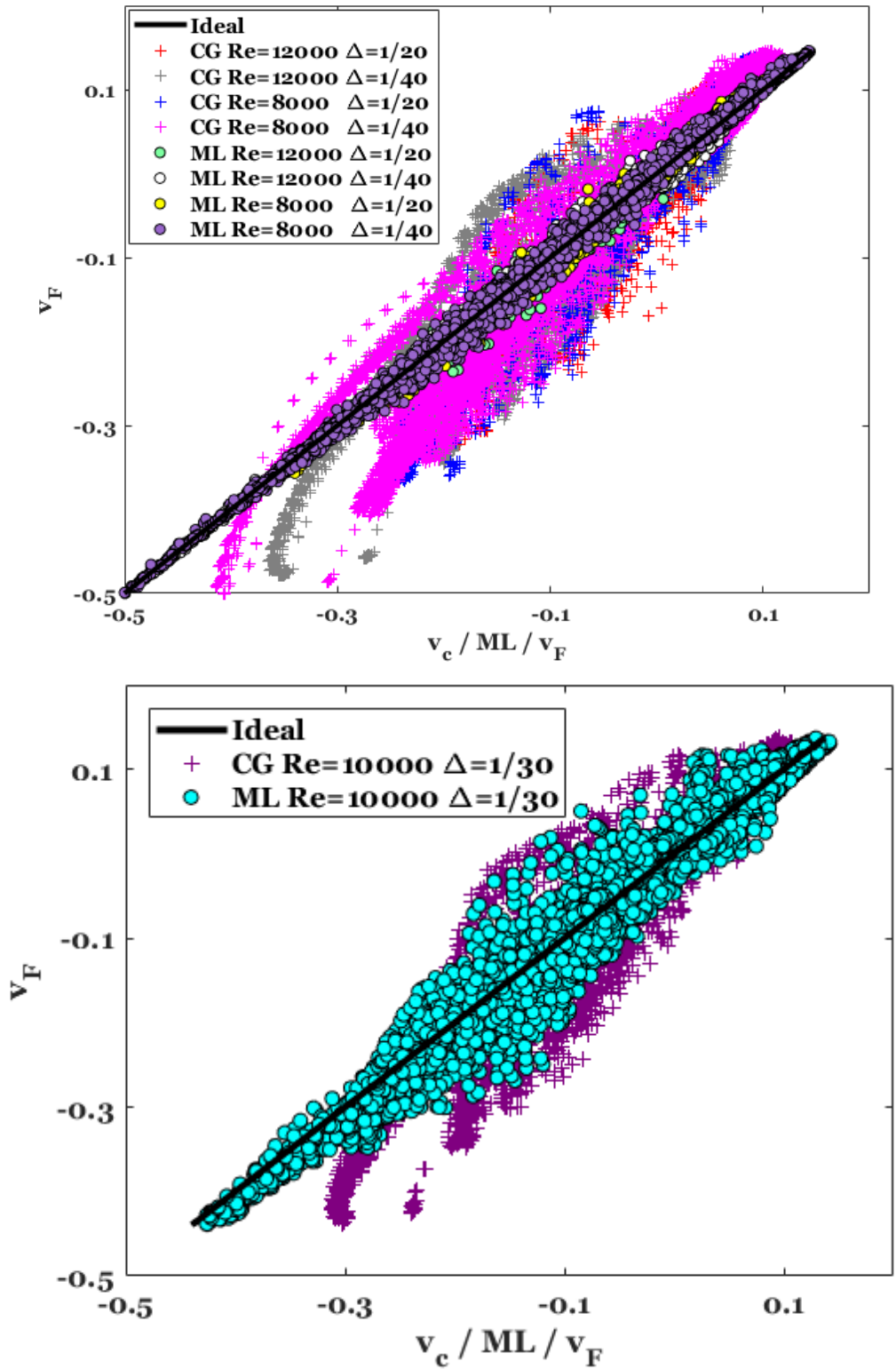

*Fig. 20. Scenario VII for* $U_y$ *(by RFR)*. Training data (above) and testing data (below).

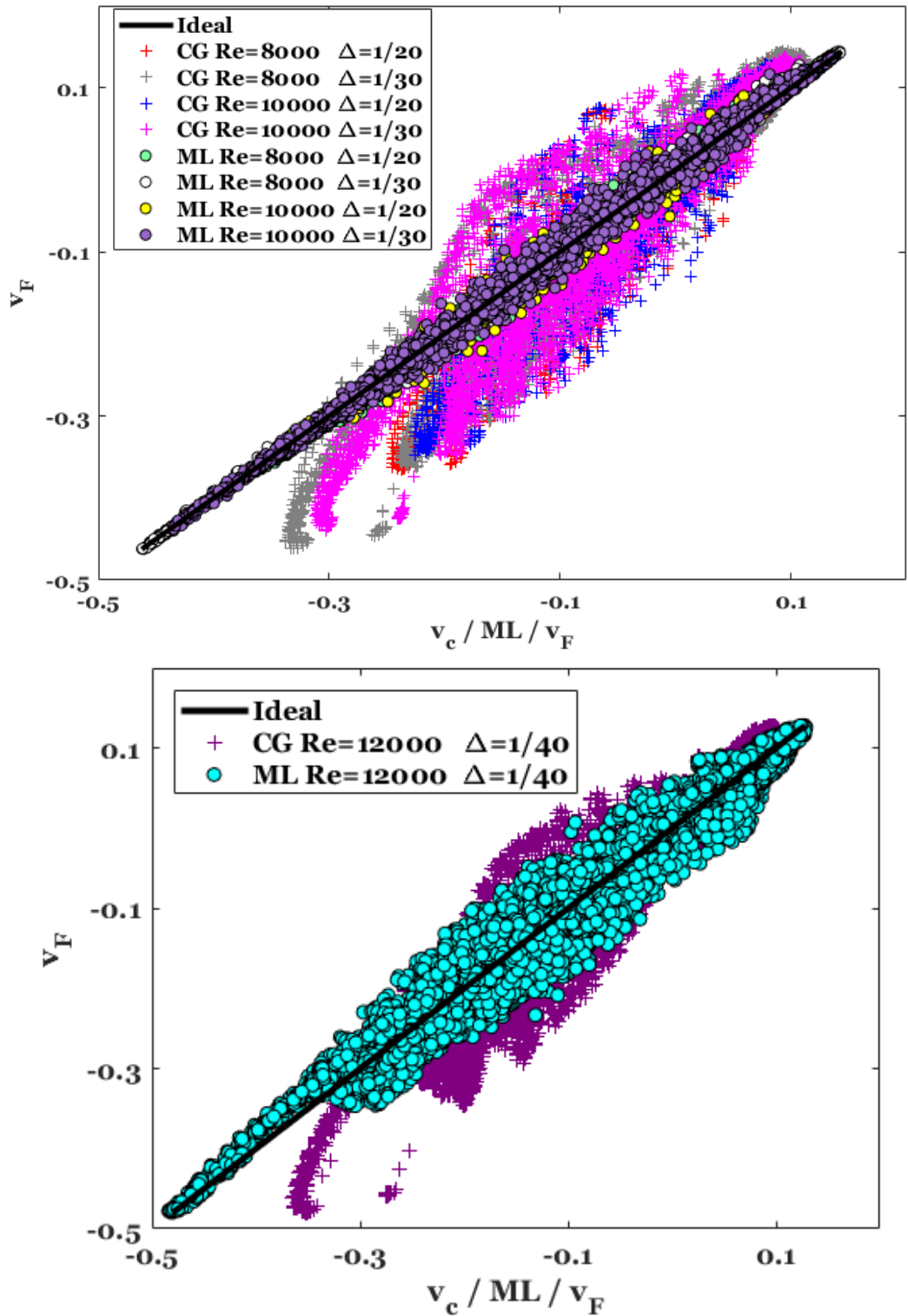

Fig. 21. Scenario VIII for $U_y$ (by RFR). Training data (above) and testing data (below).

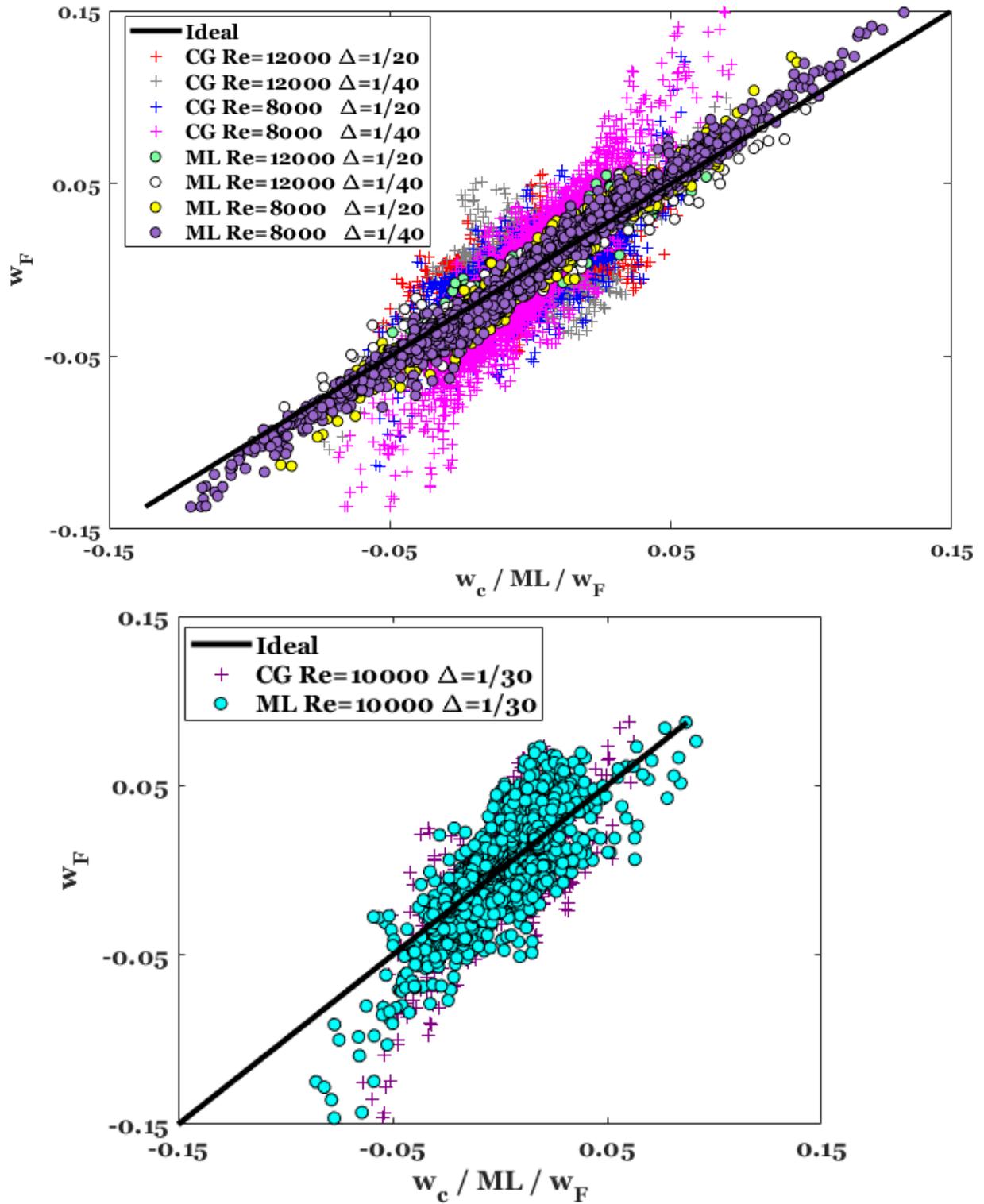

Fig. 22. Scenario VII for $U_z$ (by RFR). Training data (above) and testing data (below).

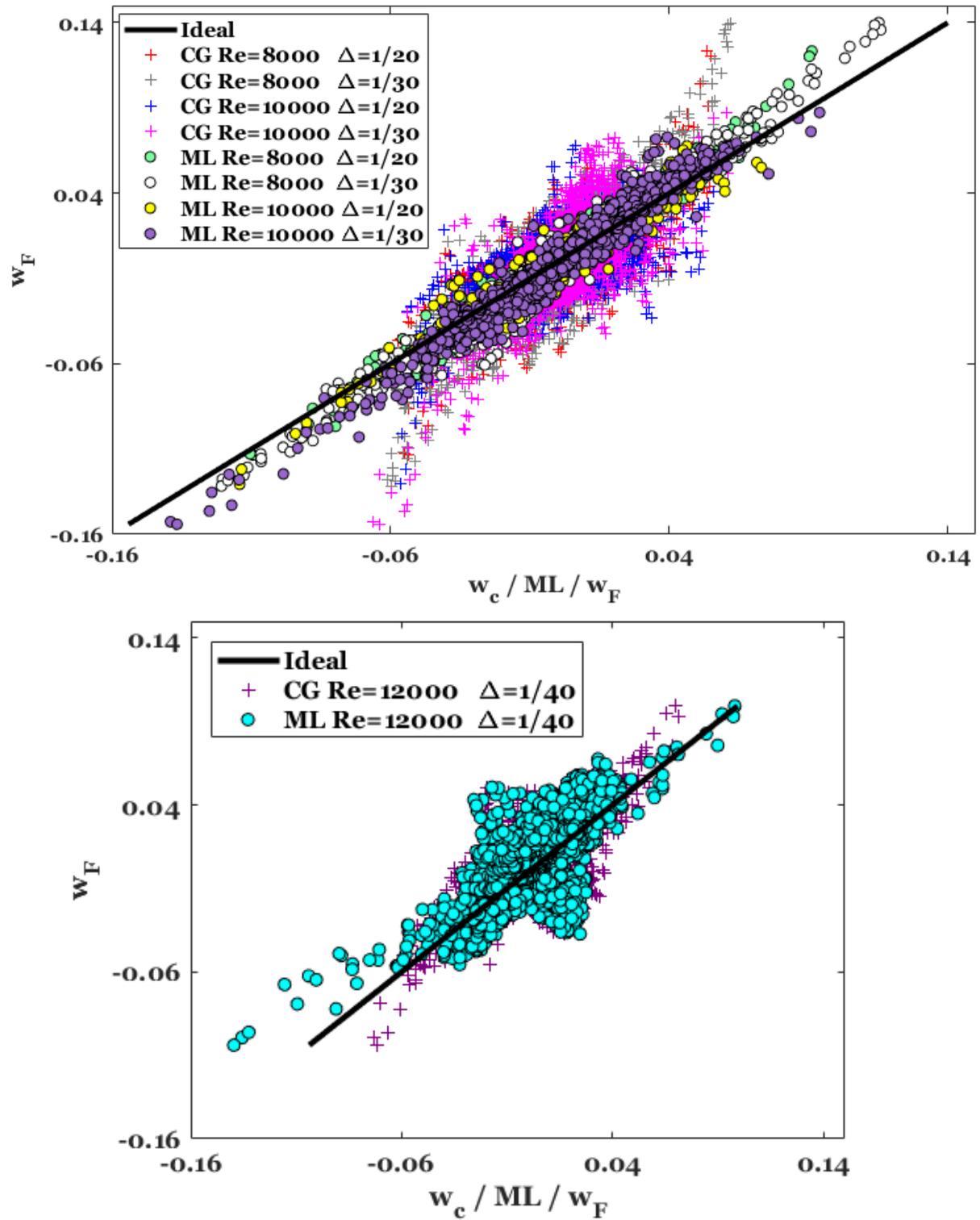

Fig. 23. Scenario VIII for $U_z$ (by RFR). Training data (above) and testing data (below).

Table 3. A comparison between different scenarios in terms of ML error and computational expense.

| Scenario | ML | Training data | | | Testing data | | Variable of interest |
|---|---|---|---|---|---|---|---|
| | | Mean ($E_{ML}$) | Max ($E_{ML}$) | Computational time (minutes) | Mean ($E_{ML}$) | Max ($E_{ML}$) | |
| I | ANN | 0.3 | 3.5 | 12.9 | 0.4 | 3.4 | $U_x$ |
| I | RFR | 0.06 | 1.5 | 11.1 | 0.3 | 2.8 | |
| II | ANN | 0.3 | 7.6 | 13 | 0.4 | 7 | |
| II | RFR | 0.06 | 1 | 11.2 | 0.4 | 2.7 | |
| III | | 0.07 | 1.3 | 11.3 | 0.4 | 2.9 | |
| IV | | 0.05 | 2 | 6.4 | 0.3 | 3 | |
| V | | 0.05 | 1.1 | 3 | 0.4 | 4 | |
| VI | | 0.05 | 1.9 | 8.2 | 0.4 | 2 | |
| VII | | 0.06 | 1.5 | 22 | 0.3 | 4.6 | |
| VIII | | 0.06 | 1.2 | 13.3 | 0.3 | 3.3 | |
| IX | | 0.07 | 1.8 | 32.3 | 0.4 | 2.3 | |
| VII | | 0.06 | 2.2 | 14.9 | 0.3 | 4.9 | $U_y$ |
| VIII | | 0.07 | 1.9 | 6.3 | 0.3 | 4.6 | |
| VII | | 0.1 | 5.3 | 17.2 | 0.6 | 10 | $U_z$ |
| VIII | | 0.1 | 4.6 | 6 | 0.7 | 13.1 | |

Table 4. Data convergence study.

| # | Testing flows | Training flows | Testing data mean $E_{ML}$ | Testing data maximum $E_{ML}$ |
|---|---|---|---|---|
| I | $Re = 12000,$ $\Delta = 1/30m$ | $Re = 6000, \Delta = 1/30m$ | 0.4 | 3.21 |
| | | $Re = 6000, \Delta = 1/30m$ $Re = 8000, \Delta = 1/30m$ | 0.36 | 2.78 |
| | | $Re = 6000, \Delta = 1/30m$ $Re = 8000, \Delta = 1/30m$ $Re = 10000, \Delta = 1/30m$ | 0.35 | 2.71 |
| II | $Re = 12000,$ $\Delta = 1/30m$ | $Re = 10000, \Delta = 1/30m$ | 0.4 | 2.67 |
| | | $Re = 10000, \Delta = 1/30m$ $Re = 8000, \Delta = 1/30m$ | 0.36 | 2.63 |
| | | $Re = 10000, \Delta = 1/30m$ $Re = 8000, \Delta = 1/30m$ $Re = 6000, \Delta = 1/30m$ | 0.35 | 2.71 |
| III | $Re = 12000,$ $\Delta = 1/40m$ | $Re = 12000, \Delta = 1/20m$ | 0.5 | 4.63 |
| | | $Re = 12000, \Delta = 1/20m$ $Re = 12000, \Delta = 1/30m$ | 0.38 | 4 |
| IV | $Re = 12000,$ $\Delta = 1/40m$ | $Re = 12000, \Delta = 1/30m$ | 0.39 | 3.89 |
| | | $Re = 12000, \Delta = 1/30m$ $Re = 12000, \Delta = 1/20m$ | 0.38 | 4 |